\documentclass[twocolumn]{aastex63}

\usepackage{CJK}
\usepackage{amsmath}

\usepackage{natbib}
\usepackage{graphicx}
\usepackage{amsmath}
\usepackage{booktabs}
\usepackage{longtable}
\usepackage{xspace}
\usepackage{hyperref}
\usepackage[T1]{fontenc}
\usepackage{lipsum}
\usepackage{comment}
\usepackage{xcolor}

\received{}
\revised{}
\accepted{}

\shorttitle{Long-period Giant Planet in Kepler-129 }
\shortauthors{Zhang et al. 2021}

\graphicspath{{./}{figures/}}

\begin{document}
\begin{CJK*}{UTF8}{gbsn}
\title{Long Period Jovian Tilts the Orbits of Two sub-Neptunes Relative to Stellar Spin Axis in Kepler-129}

\correspondingauthor{Jingwen Zhang}
\email{jingwen7@hawaii.edu}

\affiliation{Institute for Astronomy, University of Hawai`i, 2680 Woodlawn Drive, Honolulu, HI 96822, USA}

\author[0000-0002-2696-2406]{Jingwen Zhang（张婧雯）}
\affiliation{Institute for Astronomy, University of Hawai`i, 2680 Woodlawn Drive, Honolulu, HI 96822, USA}

\author[0000-0002-3725-3058]{Lauren M. Weiss}
\affiliation{Institute for Astronomy, University of Hawai`i, 2680 Woodlawn Drive, Honolulu, HI 96822, USA}

\author[0000-0001-8832-4488]{Daniel Huber}
\affiliation{Institute for Astronomy, University of Hawai`i, 2680 Woodlawn Drive, Honolulu, HI 96822, USA}

\author[0000-0002-3199-2888]{Sarah Blunt}
\altaffiliation{NSF Graduate Research Fellow}
\affiliation{Department of Astronomy, California Institute of Technology, Pasadena, CA 91125, USA}

\author[0000-0003-1125-2564]{Ashley Chontos}
\altaffiliation{NSF Graduate Research Fellow}
\affiliation{Institute for Astronomy, University of Hawai`i, 2680 Woodlawn Drive, Honolulu, HI 96822, USA}

\author[0000-0003-3504-5316]{Benjamin J. Fulton}
\affiliation{NASA Exoplanet Science Institute / Caltech-IPAC}

\author[0000-0003-4976-9980]{Samuel Grunblatt}
\affiliation{American Museum of Natural History, 200 Central Park West, Manhattan, NY 10024, USA}
\affiliation{Center for Computational Astrophysics, Flatiron Institute, 162 5$^\text{th}$ Avenue, Manhattan, NY 10010, USA}

\author[0000-0001-8638-0320]{Andrew W. Howard}
\affiliation{Department of Astronomy, California Institute of Technology, Pasadena, CA 91125, USA}

\author[0000-0002-0531-1073]{Howard Isaacson}
\affiliation{501 Campbell Hall, University of California at Berkeley, Berkeley, CA 94720, USA}
\affiliation{Centre for Astrophysics, University of Southern Queensland, Toowoomba, QLD, Australia}

\author[0000-0002-6115-4359]{Molly R.\ Kosiarek}
\altaffiliation{NSF Graduate Research Fellow}
\affiliation{Department of Astronomy and Astrophysics, University of California, Santa Cruz, CA 95064, USA}

\author[0000-0003-0967-2893]{Erik A. Petigura}
\affiliation{Department of Physics and Astronomy, University of California Los Angeles, Los Angeles, CA 90095, USA}

\author[0000-0001-8391-5182]{Lee J.\ Rosenthal}
\affiliation{Department of Astronomy, California Institute of Technology, Pasadena, CA 91125, USA}

\author[0000-0003-3856-3143]{Ryan A. Rubenzahl}
\altaffiliation{NSF Graduate Research Fellow}
\affiliation{Department of Astronomy, California Institute of Technology, Pasadena, CA 91125, USA}

\begin{abstract}
We present the discovery of Kepler-129 d ($P_{d}=7.2^{+0.4}_{-0.3}$ yr, $m\sin i_{d}=8.3^{+1.1}_{-0.7}\  \rm M_{Jup}$, $ e_{d}=0.15^{+0.07}_{-0.05} $) based on six years of radial velocity (RV) observations from Keck/HIRES. Kepler-129 also hosts two transiting sub-Neptunes: Kepler-129 b ($P_{b}=15.79$ days, $r_{b}=2.40\pm{0.04}\ \rm{R_{\earth}}$) and Kepler-129 c ($P_{c}=82.20$ days, $r_{c}=2.52\pm{0.07}\ \rm{R_{\earth}}$) for which we measure masses of $m_{b}<20\ \rm{M_{\earth}}$ and $m_{c}=43^{+13}_{-12}\ \rm{M_{\earth}}$. Kepler-129 is an hierarchical system consisting of two tightly-packed inner planets and an external companion whose mass is close to the deuterium burning limit. In such a system, two inner planets precess around the orbital normal of the outer companion, causing their inclinations to oscillate with time. Based on an asteroseismic analysis of \textit{Kepler} data, we find tentative evidence that Kepler-129 b and c are misaligned with stellar spin axis by $\gtrsim 38^{\circ}$, which could be torqued by Kepler-129 d if it is inclined by $\gtrsim19\degr$ relative to inner planets. Using N-body simulations, we provide additional constraints on the mutual inclination between Kepler-129 d and inner planets by estimating the fraction of time during which two inner planets both transit. The probability that two planets both transit decreases as their misalignment with Kepler-129 d increases. We also find a more massive Kepler-129 c enables the two inner planets to become strongly coupled and more resistant to perturbations from Kepler-129 d.  The unusually high mass of Kepler-129 c provides a valuable benchmark for both planetary dynamics and interior structure, since the best-fit mass is consistent with this $\rm{2.5\ R_{\oplus}}$ planet having a rocky surface.

\end{abstract}


\section{Introduction} \label{sec:intro}
Our solar system is a multi-planet system, in which planets orbit in the solar equatorial plane with only a few degrees dispersion. The alignment between solar spin and planetary orbital axes is considered to result from the protoplanetary disk where these planets formed and has been maintained throughout the history of the solar system \citep{Kant1775,laplace1796}. On the contrary, spin-orbit misalignment has been found in dozens of exoplanet systems \citep[e.g.][]{Winn2010,Huber2013, Yee2018, Kamiaka2019, Bourrier2018,Rubenzahl_2021}, suggesting a different formation pathway or the occurrence of dynamical events in those systems. Measuring the spin-orbit angle therefore helps to understand the formation and evolution of planetary systems. The measurement of spin-orbit misalignment can be achieved by the Rossiter-McLaughlin (RM) effect which is based on monitoring the sequential distortion in stellar radial velocity (RV) during a planetary transit \citep{R1924, M1924, Winn2005}. Similarly, when the stellar lines are sufficiently broadened from rotation, it is possible to spectrally resolve the "shadow" of the planet occulting a specific part of the line in what is called the Doppler shadow technique \citep{Cameron2010}. The shape of distorted RVs or rotationally broadened spectral lines depends on the spin-orbit angle projected in the sky-plane. Because the sizes of RM effect and Doppler shadow scale with squared planet radius, they have mostly been used for hot Jupiters. It is much more difficult to apply this technique for super-Earth sized planets.  

Another way to probe the spin-orbit misalignment is through asteroseismology \citep{Gizon2003}. The stellar rotation induces splittings in oscillation modes, which can be used to measure the direction of the stellar spin axis relative to the line of sight. When the host star has transiting planets with nearly edge-on orbits, the difference in inclination between the star and planetary orbits can be obtained \citep{chaplin_etal2013}. Because the asteroseismic method depends on the stellar parameters but not on the planet size, it can be used to measure the spin-orbit angle in systems with smaller planets. A small but growing number of close-in transiting super-Earths and sub-Neptunes have been observed to possess spin-orbit misalignment \citep[e.g.][]{Huber2013, Kamiaka2019, Bourrier2018}. For example, Kepler-56 b and c are two small transiting planets that orbit in a plane inclined with respect to the stellar equator by $\sim\ 37^{\circ}$ \citep{Huber2013}. 

Several theories have been proposed to explain spin-orbit misalignments in systems with close-in super-Earths or sub-Neptunes. One recently proposed mechanism is that close-in small planets are tilted by a rapidly rotating young star  which is highly oblate \citep{Spalding2014, Spalding2020}. If the star is misaligned with respect to its protoplanetary disk by some mechanism such as torquing from another star or star disk magnetic torques, it might tilt the orbits of close-in planets or even excite mutual inclinations between the planets. Another possible mechanism is that planets formed in warped protoplanetary disks and intrinsically possess mutual inclination relative to each other, although observations of disk warps are currently limited \citep{zanazzi2018}.

Both of the above scenarios occur in the early age of the system. Alternatively, close-in super-Earths or sub-Neptunes could be aligned with stellar spin when they formed, but are later tilted out of alignment by an inclined outer giant planet. This scenario requires a non-zero mutual inclination between the inner small planets and outer giant planet(s), which could be caused by dynamical events such as planet-planet scattering \citep{chatterjee2008}. An example in favor of this scenario is HAT-P-11, which hosts a transiting planet with a projected spin-orbit angle of $\sim100^{\circ}$ \citep{Winn2010,Hirano2011}. \citet{Yee2018} later discovered an eccentric outer giant planet ($e\approx0.6$) in HAT-P-11 and proposed that the misalignment could be caused by nodal precession of the inner orbit around that of the outer giant's orbit. Using {\it Gaia} DR2 and {\it Hipparcos} astrometry, \citet{Xuan2020} measured a mutual inclination of $>54^{\circ}$ ($1\sigma$) between the two planets in HAT-P-11, which supports this picture. In addition, the transiting super Earth and giant planet in $\pi$ Men were found to be mutually inclined by $\sim\ 50^{\circ}$ \citep{Xuan2020,Damasso2020,DeRosa2020}. Soon after, the super Earth in $\pi$ Men was found to be moderately misaligned with the stellar spin axis by $\sim30^{\circ}$ with the Doppler Shadow technique \citep{Kunovac2020}. 

So far, direct measurements of mutual inclination between inner small planet and outer giant planets are still limited to a few systems. Indirect methods can be used to set a constraint on the mutual inclination. For example, based on the number of transiting giant planets in {\it Kepler} systems that also host transiting super Earths, \citet{Masuda2020} estimated the average mutual inclination between inner super Earths and outer giant planets to be around $11^{\circ}.8^{+12.7}_{-5.5}$ ($1\sigma$ confidence). In addition, an inclined outer perturber may increase mutual inclinations between the inner planets or even destabilize their orbits \citep{Becker2017,DongandPu2017,Huang2017,PL2018,denham_hidden_2019}. Thus, if a system hosts two or more transiting planets, the double transit probability can provide an upper boundary for the mutual inclination between the inner small planets and outer giant planets \citep[e.g.][]{Becker2017}. 

Kepler-129 (KOI-275) hosts two transiting planets, Kepler-129 b and c, which are sub-Neptunes with orbital periods of 15.79 days and 82.20 days, respectively \citep{Rowe2014,Van2015}. We report the discovery of a long-period giant planet (hereafter Kepler-129 d) based on RV observations with the Keck/HIRES spectrograph and study the interaction between inner small planets and the outer giant planet in this system. 

\section{Observations} \label{sec:obs}
\subsection{System Parameters} \label{subsec:starpara}
Kepler-129 is a sub-giant G4V star at an age of $6.43^{+0.64}_{-0.61}$ Gyr at a distance of $402.73^{+12.28}_{-12.43}\ \rm{pc}$ \citep{SA2015}. The measured mass is  $1.178^{+0.021}_{-0.030}\ M_{\odot}$, radius is $1.653^{+0.009}_{-0.012}\ R_{\odot}$ and [Fe/H] is $0.29\pm{0.10}$ \citep{FP18,SA2015}. Its projected rotation velocity $v\sin i$ determined spectroscopically is $2.13\pm{1.0}\ km\ s^{-1}$\citep{Petigura_thesis}. The system has two transiting planets, Kepler-129 b and c, which are two sub-Neptunes ($r_{b}=2.40\pm{0.04}\ r_{\earth}, r_{c}=2.52\pm{0.07}\ r_{\earth}$) with orbital periods of 15.79 days and 82.20 days and eccentricities of $0.01^{+0.24}_{-0.01}$ and $0.20^{+0.15}_{-0.20}$ \citep{Rowe2014,Van2015}. The stellar and planetary parameters are given in Table~\ref{tab:star-compare}.

\subsection{Keck/HIRES Radial Velocities}\label{subsec:Hobs}
We collected the RV data for Kepler-129 from 2014 to 2021 using the High Resolution Echelle Spectrometer (HIRES, R $\sim$60000) \citep{Vogt1994} at the W.M. Keck Observatory. The observations are part of the Kepler Giant Planet Survey, which aims to search for long-period giant planets around 60 Kepler stars with HIRES (Weiss et al. in prep). The observation setup is the same as that used by the California Planet Search \citep{Howard2010}. We used the C2 decker ($0.86^{\prime\prime} \times 14^{\prime\prime}$) to subtract the contaminating light from the sky background. The wavelength calibration was done with a iodine gas cell in the light path. A iodine-free template spectrum bracketed by observations of rapidly rotating B-type stars was used to deconvolve the stellar spectrum from the spectrograph PSF. We then forward model the spectra taken with the iodine cell using the deconvolved template spectra. The wavelength scale, the instrumental profile and the RV in each of the $\sim 700$ segments of 80 pixels were solved simultaneously \citep{Howard2010}. Our Keck-HIRES RVs are presented in Table~\ref{tab:rvs}.

\begin{deluxetable}{lcccc}




\tablecaption{Kepler-129 RVs\label{tab:rvs}
}

\tablehead{\colhead{Time} & \colhead{RV} & \colhead{$\sigma_\mathrm{RV}$} & \colhead{$S_{HK}$}& \colhead{Inst} \\ 
\colhead{(BJD - 2450000)} & \colhead{(m/s)} & \colhead{(m/s)} &\colhead{}& \colhead{} } 

\startdata
6912.968 & -52.65 & 3.00 & 0.1256 &HIRES \\
8263.903 & 67.70 & 2.02 & 0.1320 &HIRES \\
8302.092 & 77.14 & 2.54 &0.1292 &HIRES \\
8329.813 & 57.54 & 2.14 & 0.1317 &HIRES \\
8337.042 & 49.42 & 2.19 &0.1301 &HIRES \\
...         & ... & ... & ... &... \\
\enddata



\tablecomments{Times are in BJD - 2450000.0. The RV uncertainties do not include RV jitter.  The full table is available in machine readable form.  The first few lines are shown here for content and format.}

\end{deluxetable}
\subsection{Kepler Photometry}\label{subsec:KP}
We used Kepler short cadence time series ($\Delta$t$\sim$58.85s) to detect the solar-like oscillations of Kepler-129 with a typical period of a few minutes. The available short cadence data for Kepler-129 consists of two parts, one spans from Q6.1 to Q7.3 (Jun. 24 2010-Dec. 22 2010 ), and another was collected during Q17.1 and Q17.2 (Apr. 09 2013-May. 11 2013). We reduced the data with the \textit{lightkurve} \citep{LK2018} package. The frequency power spectra were computed using a Lomb-Scargle periodogram \citep{Scargle1982}. In order to avoid the window effect due to the long gap between Q6/Q7 and Q17, we removed it from the light curves by making the Q17 timestamps consecutive with those of Q7. The process is justified since the oscillations are not phase coherent and the length of the gap is much larger than the period and mode lifetime of the oscillations \citep{Hekker2010}.

\begin{figure}
    \centering
    \includegraphics[width=\linewidth]{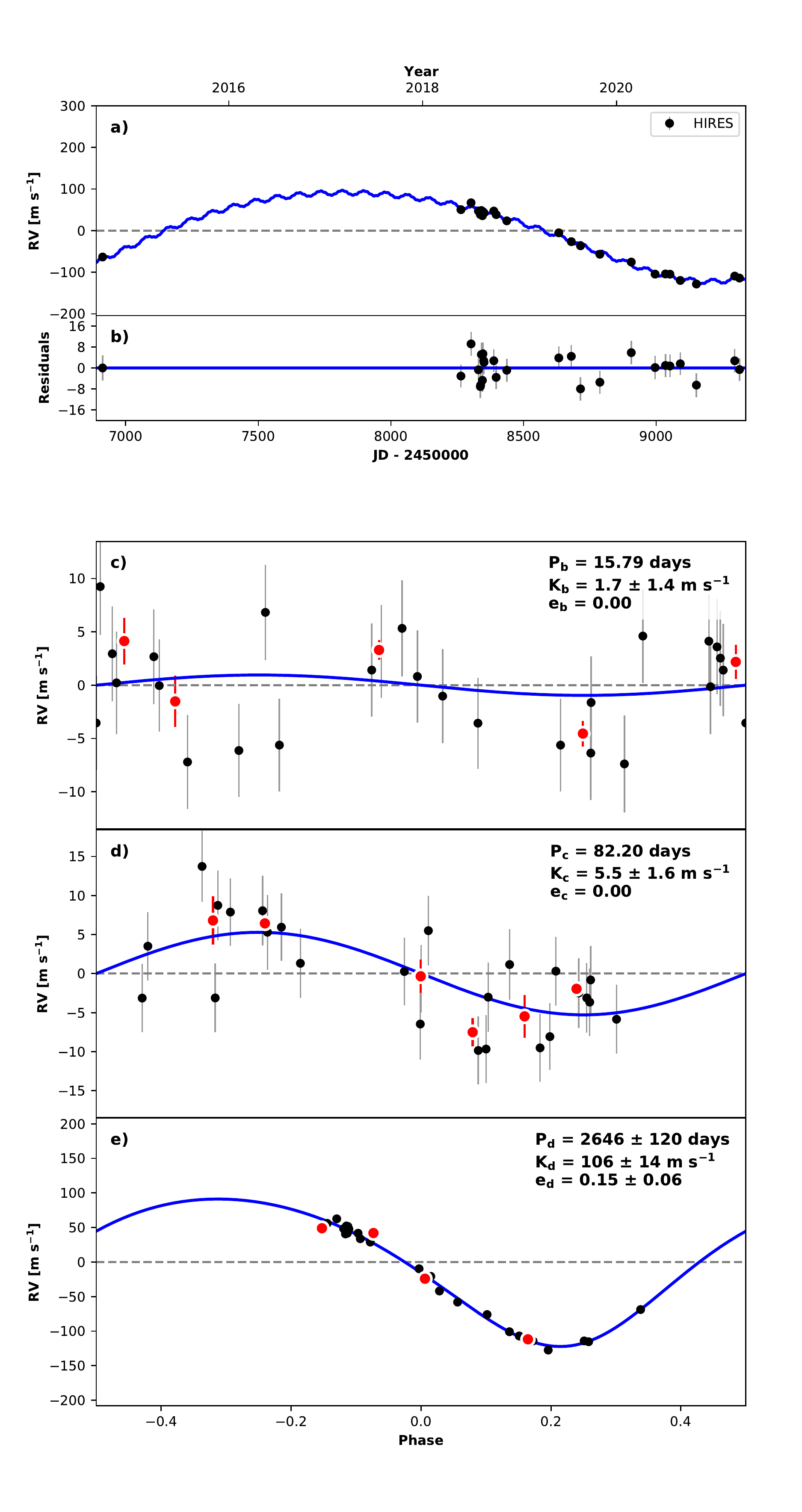}
    \caption{Best-fit 3-planet Keplerian orbital model for Kepler129. a: Kepler-129 RVs with errors (black) and their best fit model (blue) as a function of time. b: the residuals. c$\sim$e: RV data and models for each planet phase-folded at the best-fit orbital period with all other planets' signals removed. The orbital periods, time of conjunction and eccentricities for planet b and c are fixed since they are well constrained from transit observations. }
    \label{fig:figure1}
\end{figure}

\section{Keplerian Fit}\label{sec:Kfit}
\subsection{Maximum Likelihood Fitting}\label{subsec:xfit}
The RVs of Kepler-129 reveal a long term variation from a planetary companion, and the single data point collected in 2014 provides a constraint that the long period is approximately 3000 days. We used \textit{RadVel} \citep{Fulton2018}, a Keplerian multi-planet RV fitting package, to obtain orbital properties for all three planets. Keplerian orbits are fitted with five orbital elements $K$, $lnP$, $T_{conj}$, $\sqrt{e}sin\omega$,$\sqrt{e}cos\omega$, where $K$ is the RV semi-amplitude, $P$ is the orbital period, $T_{conj}$ is the time of conjunction, $e$ is the eccentricity, and $\omega$ is the argument of pericenter. In addition, a HIRES RV zeropoint $\gamma$ and a RV jitter term $\sigma$ are fitted in the models. We fix the orbital periods and times of conjunction of the two inner planets Kepler-129 b and c since they are well determined by Kepler observations \citep{Van2015}. We also set their eccentricity and argument of pericenter as 0 to simplify the fitting as they are likely to have low eccentricities \citep{Van2015}. Therefore, for the two inner planets, the only free parameters are their $K$ amplitudes. For the outer giant planet, we allow all five of its orbital parameters to vary. In addition, we set bounds on $0<e<1$, $K>0$, $0<\sigma<10$ for all planets. The set of orbital parameters was determined based on minimum $\chi^{2}$ fitting. Figure~\ref{fig:figure1} shows the best fit Keplerian solution. 

\subsection{Parameter uncertainties with MCMC} \label{subsec:br}
We performed the Markov Chain Monte Carlo (MCMC) exploration with emcee \citep{FM2013} to estimate parameter credible levels. Our MCMC analysis used 50 walkers and ran for $\sim 10^4$ steps per walker, achieving a maximum Gelman-Rubin(GR) statistic of 1.005. We derived the (minimum) planet mass from RV amplitudes ($m_{b}< 20 \ \rm M_{\earth}\ (95\%)$, $ m_{c}=43^{+13}_{-12}\ \rm M_{\earth}$, $ m\sin i_d=8.3^{+1.1}_{-0.7}\ \rm M_{Jup}$). The derived planetary parameters are given in Table~\ref{tab:star-compare}.
\begin{figure}
    \centering
    \includegraphics[width=\linewidth]{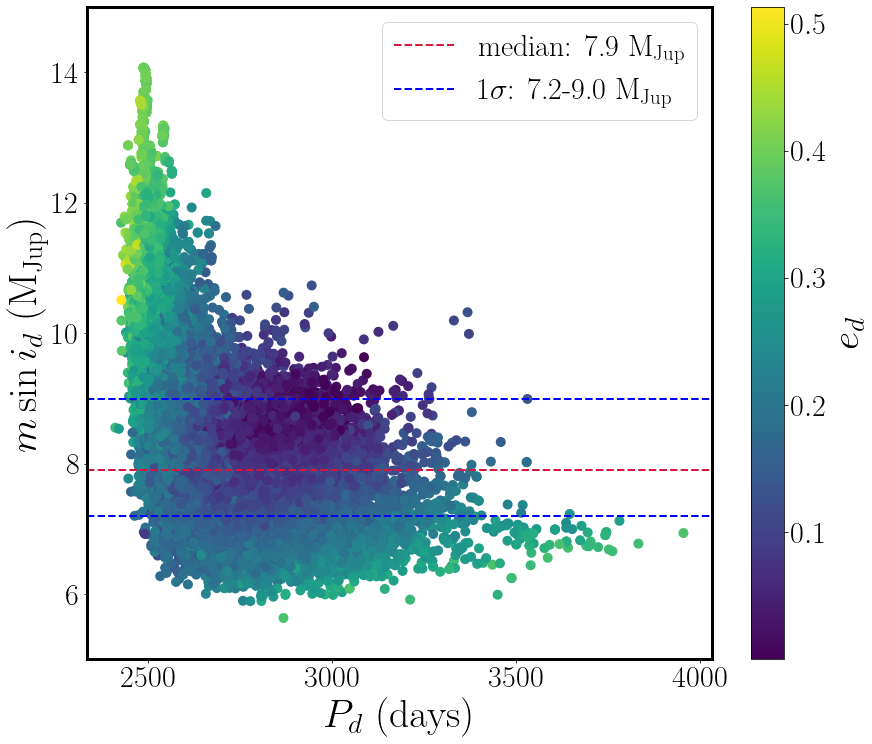}
    \caption{The minimum mass of Kepler-129 d as a function of the orbital period colored by eccentricity.   The red dashed line corresponds to median value of $msini$, and blue dashed lines indicate the $1\sigma$ confidence interval. Kepler-129 d is a massive long-period giant planet with low eccentricity, whose minimum mass distribution has a  tail beyond the traditional boundary between planets and brown dwarfs based on the deuterium burning limit (13 $\rm M_{Jup}$).}
    \label{fig:figmass}
\end{figure}

Figure~\ref{fig:figmass} shows that the minimum mass distribution of Kepler-129 d has a tail beyond 13 $M_{Jup}$, which is the traditional boundary between planet and brown dwarf based on deuterium burning limit\citep{GandG1973}. Given that the $\sin i_{d}$ is unknown, the true mass of Kepler-129 d could be larger, possibly pushing it into brown dwarf regime under this definition. On the other hand, several studies argue for the formation-based definition that planets form through core accretion and brown dwarfs form due to gravitational instability\citep{Schlaufman2018}. The formation channels predict two patterns: objects formed through core accretion are preferentially found around metal-rich stars and with low eccentricity, whereas those formed through gravitational collapse occur with equal efficiency independent of stellar metal abundance and tend to have larger eccentricity\citep{Schlaufman2018,Bowler2020}. Our results of low eccentricity ($e_{d}=0.15^{+0.07}_{-0.05}$) of Kepler-129 d, as well as high metal abundance of the star ([Fe/H]$\sim 0.26$), is consistent with a planet definition, although we can not rule out the possibility of the brown dwarf. For the sake of simplicity, we will refer to Kepler-129 d as a giant planet in this paper.

\begin{figure}
    \centering
    \includegraphics[width=\linewidth]{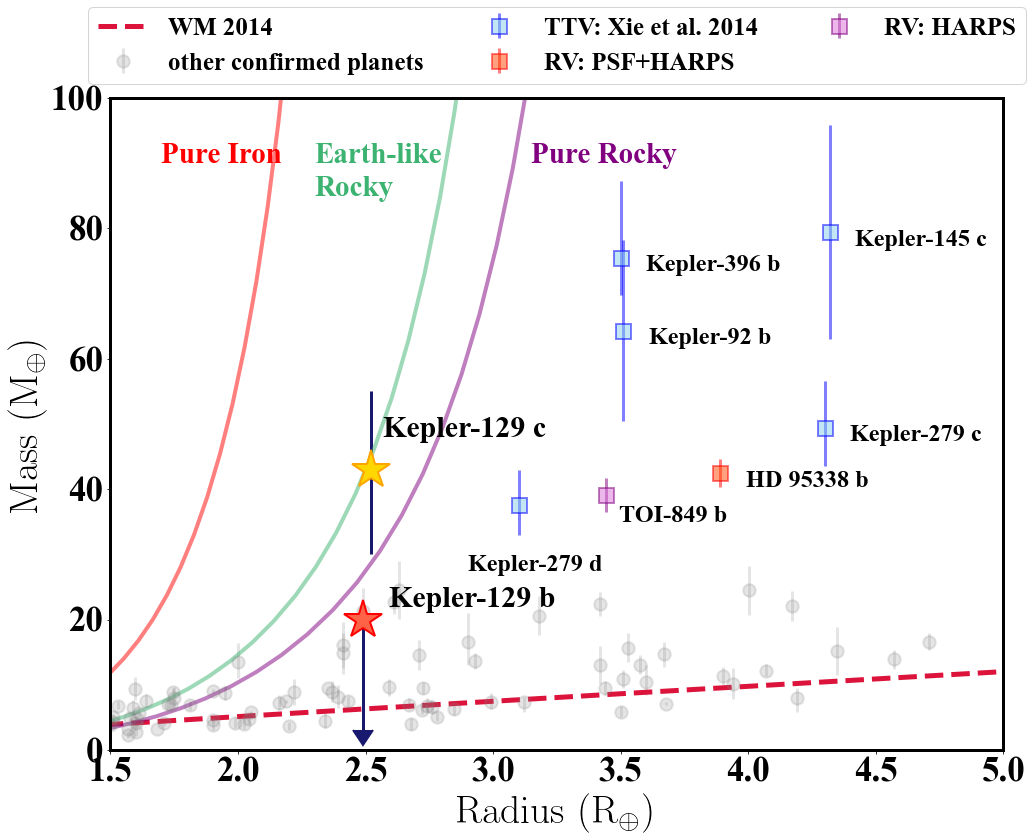}
    \caption{Mass vs. radius for Kepler-129 b and c in comparison to other confirmed planets from NASA Exoplanet Archive. The plotted planets have orbital periods smaller than 100 days, radius between 1.5 $R_{\oplus}$ and 5 $R_{\oplus}$, and $<25\%$ mass uncertainty. The solid lines are planet composition lines from \cite{zeng2008}. The red dashed line corresponds to the mass radius relation given by \cite{WM2014}. The mass range of Kepler-129 b overlaps with the \cite{WM2014} relation, which suggests the possibility of Kepler-129 b having a H/He envelope. Kepler-129 c has a unique high mass, indicating it may be a rocky core without atmosphere. }
    \label{fig:mass_radii}
\end{figure}

\begin{deluxetable*}{p{5cm}ccc}
\tablecaption{Stellar and Planetary Properties}
\label{tab:star-compare}

\tablehead{\colhead{Parameter} & \colhead{Credible Interval} &\colhead{Units} & \colhead{Reference} }
\startdata
\sidehead{\bf{Stellar Parameters}}
$T_{\rm eff}$      & $5770\pm83$   & K  & A \\
$M_{\star}$  & $1.178^{+0.021}_{-0.030}$  & $M_{\odot}$ & A \\
$R_{\star}$  & $1.653^{+0.009}_{-0.012}$  &  $R_{\odot}$ & A\\
 $\rm [Fe/H]$    & $0.29\pm0.10$  &   dex & A  \\
 $v\sin i$    &    $2.13\pm{1}$    & $km\ s^{-1}$  & D \\
Age      & $6.43^{+0.64}_{-0.61}$       & Gyr & A\\
$k_{2}$ & $0.001$   &     & C\\
$C$ & $0.05$& $M_{\star}R_{\star}^{2}$        & C\\
Stellar Inclination      & $52^{+10}_{-13}$     & deg & This work\\
\sidehead{\bf{Kepler-129 b}}
$r_{b}$     & $2.40\pm0.04$  & $R_{\oplus}$   & B     \\
$P_{b}$ &  15.79 & days & B  \\
$a_{b}$    &   $0.13$ & AU & B  \\
$T_{conjb}$ &  2454978.2      & JD  & B \\
$K_{b}$   &  $<4.5(95\%)$   &  $m\ s^{-1}$  & This work \\
$m_{b}$    &  $<20 (95\%)$   & $M_{\oplus}$  & This work        	\\
$\rho_{b}$      &  $<8.1 (95\%)$  &  $g cm^{-3}$ & This work        \\
\sidehead{\bf{Kepler-129 c}}
$r_{c}$    & $2.52\pm0.07$   & $R_{\oplus}$  & B      \\
$P_{c}$ &  82.20 & days & B  \\
$a_{c}$    &  $0.39$& AU & B   \\
$T_{conjc}$ & 2455041.8     & JD  & B \\
$K_{c}$   &  $5.5\pm{1.6}$   &  $m\ s^{-1}$  & This work \\
$m_{c} $    &  $43^{+13}_{-12}$  & $M_{\oplus}$  & This work        	\\
$\rho_{c}$      & $14.8\pm{4.3}$ &   $g\ cm^{-3}$  & This work       \\
\sidehead{\bf{Kepler-129 d}}
$P_{d}$ &  $2646^{+140}_{-94}$ & days & This work \\
$a_{d} $  &  $4.0 \pm{0.1}$ &    AU  & This work \\
$T_{conjd}$ & $2458637^{+42}_{-70}$     & JD  & This work \\
$K_{d}$   &  $106^{+17}_{-10} $  &  $m\ s^{-1}$  & This work \\
$e_{d}$   & $0.15^{+0.07}_{-0.05}  $ &   & This work\\
$w_{d}$ & $-3.0^{+0.8}_{-1.0}  $ & radians & This work\\
$m\sin i_{d}$    &  $8.3^{+1.1}_{-0.7}$ &  $\rm M_{Jup}$  & This work \\
\sidehead{\bf{Other Parameters}}
$\sigma $    &  $5.0^{+1.2}_{-1}$ &   $m\  s^{-1}$   & This work \\
$\gamma$    &  $14^{+19}_{-13}$ &  $ m\  s^{-1}$   & This work \\
$\dot{\gamma}  $    &  $\equiv 0$ &     & This work \\
$\ddot{\gamma} $     &  $\equiv 0$ &    & This work \\
\enddata
\tablecomments{  A.\citet{SA2015}; B.\citet{Van2015}; C.\citet{landin2009}. D.\cite{Petigura_thesis} $k_{2}$ is the stellar second fluid Love number and $C$ is the stellar moment of inertia along the short axis. Intervals are $68\%$ credible unless stated otherwise.}
\end{deluxetable*}

Figure~\ref{fig:mass_radii} shows masses and radii of Kepler-129 b and c in comparison to masses and radii of other sub-Neptunes from NASA Exoplanet Archive. We only selected planets with mass uncertainties $\sigma_{M}/M<25\%$. The mass range of Kepler-129 b overlaps with the predicted mass at $2.4\ \rm R_{\oplus}$ using mass radius relation in \cite{WM2014}. It likely consists of a rocky/iron core and a gaseous H/He envelope \citep{Lopez2014}. Kepler-129 c has a unique high mass, significantly ($3\sigma$) above the \cite{WM2014} mass and radius relationship, although it has nearly the same radius with Kepler-129 b. To better constrain the mass of Kepler-129 c, more RVs are needed. The high mass of Kepler-129 c indicates it may be a rocky core without atmosphere. Although Kepler-129 b and c likely formed in similar environment, they have very different masses, the reason for which is unclear but would be interesting for studies of planet composition and formation. For comparison, we identified 7 other planets with high mass in Figure~\ref{fig:mass_radii}. Five of them (blue squares), measured by \cite{Xie} using the TTV method, have relatively large uncertainties due to the degeneracy between planet mass and orbital eccentricity. The mass of other two planets, TOI-849 b (purple square) and HD 95338 b (red square), are better constrained with RV observations. TOI-849 b, with a ultra-short orbital period $<1$ day, is considered as a remnant core of a giant planet \citep{Armstrong}. HD 95338 b, with orbital period $\sim 55$ days, could be a Neptune-sized planet with a dense atmosphere \citep{D2020}. If its high mass is confirmed, Kepler-129c would be the most massive rocky planet discovered yet.

\section{Asteroseismic Analysis} \label{sec:AA}
Kepler-129 is a hierarchical system consisting of tightly packed inner planets and an external companion. The inner planetary system may be disturbed by the outer giant planet. One aspect that could shed light on the system's dynamical history is the evolution of the angle between the stellar spin axis and the total orbital angular momentum of the inner planetary system, namely the spin-orbit angle. Unless otherwise specified, we always define the spin-orbit angle with respect to the inner planetary system in this paper. \cite{Campante2016} measured the spin-orbit angle of Kepler-129 with asteroseismology, but only consider Kepler data collected in Q6/7. In this section, we present our measurements of the spin-orbit angle based on the asteroseismic analysis using Kepler data collected in Q6/7 and Q17. 

\subsection{Principles of the Method} \label{subsec:PM}

\begin{figure}
    \centering
    \includegraphics[width=\linewidth]{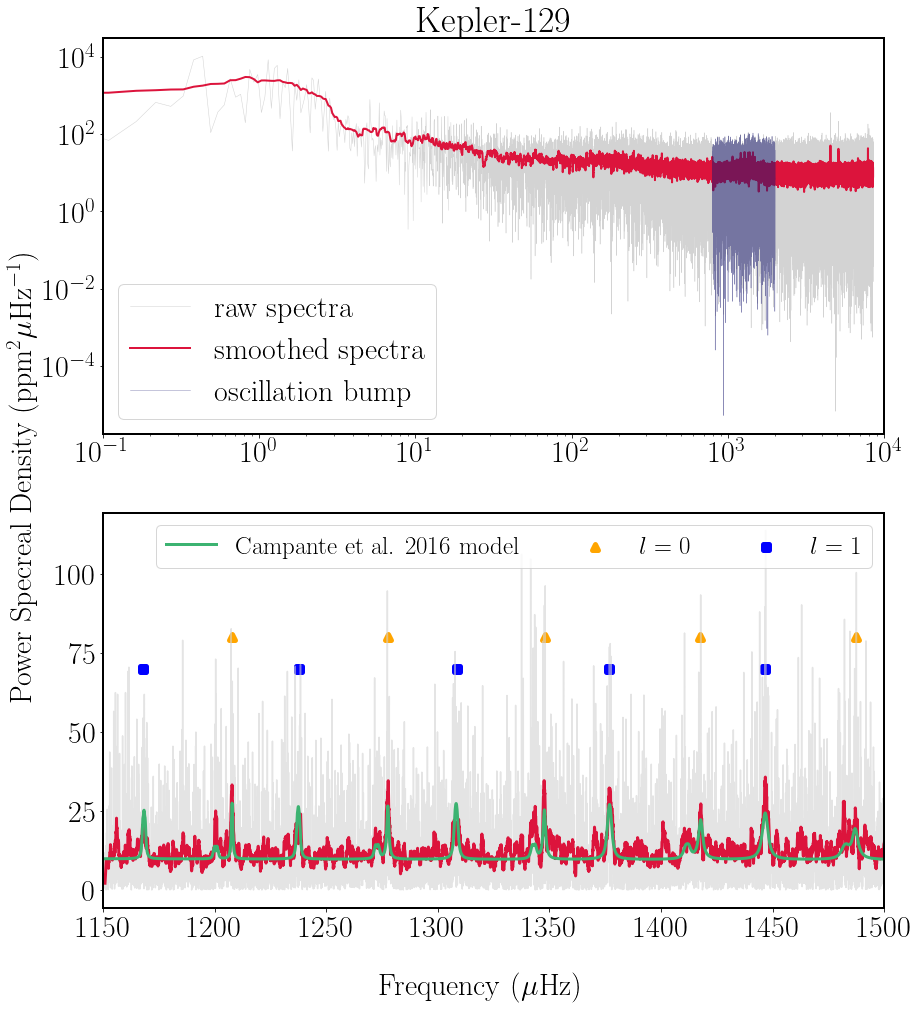}
    \caption{Power spectrum of Kepler-129. Top panel: log-scale frequency-power spectra using Lomb-Scargle periodogram. The light grey lines are original spectra, whereas red lines show the spectra after smoothing with a 1.0 $\mu$Hz filter. Dark blue region indicates the oscillation bump. Bottom panel: same as top panel, but in the frequency range of the oscillation bump. $l=0,1$ modes are marked. The green line shows the best-fitting model from \citet{Campante2016}. Kepler-129 presents clear solar-like oscillations, appearing as a pattern of evenly-spaced peaks in the Frequency-power spectra. }
    \label{fig:figure4}
\end{figure}

Solar-like oscillations are acoustic global standing waves stochastically excited and damped by near-surface convections and enable measurements of the angle between the stellar spin axis and light of sight, $i_{s}$ \citep{Gizon2003,Ballot2006,Ballot2008,Campante2011,Campante2016}. The oscillation modes, characterized by the radial order $n$, the spherical degree $l$, and the azimuthal order $m$, are typically observed in the frequency power spectrum showing a pattern of peaks with near-regular frequency separations\citep{Vandakurov1967}. Figure~\ref{fig:figure4} shows the power spectra of light curves of Kepler-129, presenting clear patterns of peaks from solar like oscillations near 1300 $\mu$Hz. The overtones of radial ($l=0$) and low-order non-radial ($l=1$) modes are detectable. 

The asteroseismic determination of $i_{s}$ is based on resolving the rotational splitting of non-radial modes in the power spectra. Rotation introduces a dependence of oscillation frequencies of non-radial modes on $m$, with prograde ($m>0$) and retrograde ($m<0$) modes having frequencies slightly higher or lower than the axisymmetric mode ($m=0$) in the observer's frame of reference \citep{Gizon2003,Chaplin2013}. For stars assumed to rotate as a solid body with angular velocity $v_{\star}$, the frequency $\nu_{nlm}$ can be expressed to first order as\citep{Ledoux1951}:
\begin{equation}
\nu_{nlm}=\nu_{nl0}+m\frac{v_{\star}}{2\pi}(1-C_{nl})\approx\nu_{nl0}+m\delta\nu_{s}
\end{equation}
where $C_{nl}$ is the dimensionless Ledoux constant to correct the effect of the Coriolis force and $C_{nl} \ll 1$ for high-order modes (large $n$) so that the rotational splitting can be given approximately by the stellar angular velocity, $\delta \nu_{s}\approx v_{*}/2\pi$.

The dependence of mode power on $m$ can be written as \citep{Dz1977,Gizon2003}:
\begin{equation}
\varepsilon_{lm}(i_{s})=\frac{(l-\mid m \mid)!}{(l+\mid m\mid)!}\left[P_{l}^{\mid m \mid}(cos(i_{s}))\right]^{2}
\end{equation}
where $P_{l}^{\mid m \mid}(x)$ are the associated Legendre functions and the sum of $\varepsilon_{lm}(i_{s})$ over m has been normalized to unity. Hence, measuring the relative power of the azimuthal components in a non-radial multiplet provides a direct estimate of the stellar inclination angle $i_{s}$. 

\begin{figure*}
    \centering
    \includegraphics[width=0.9\linewidth]{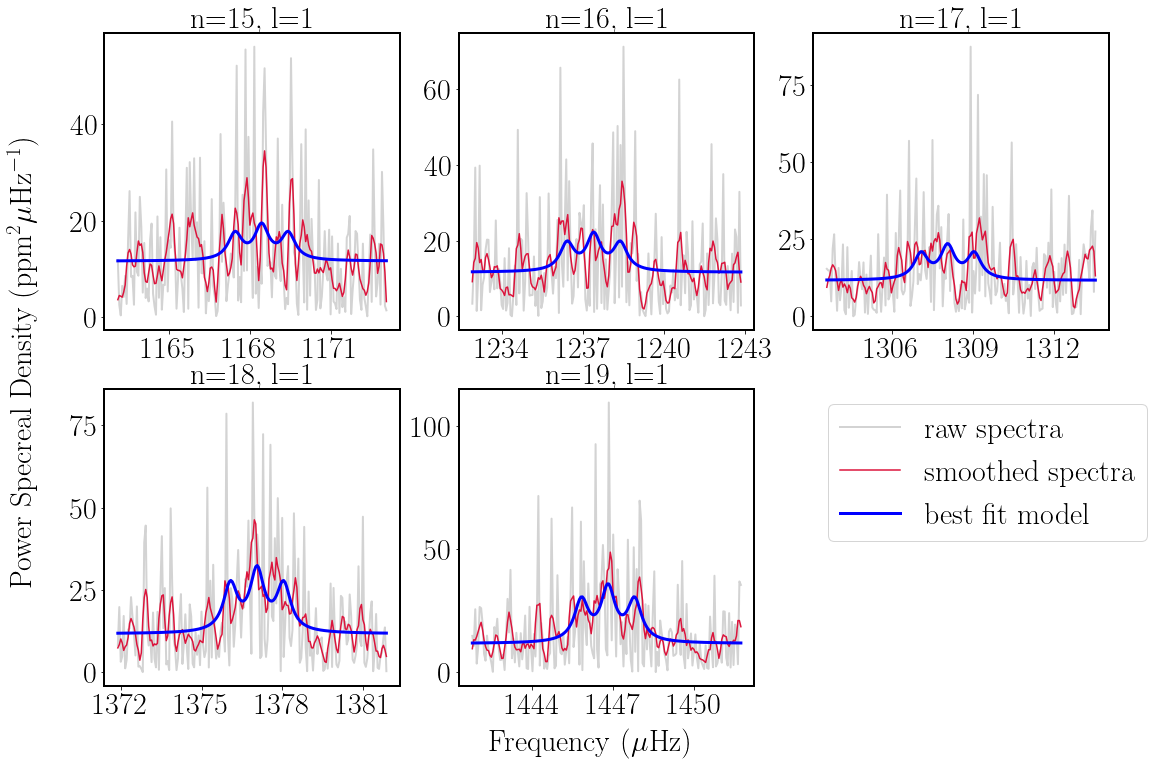}
    \caption{The rotational frequency splitting in five $l=1$ modes. Light grey lines: original spectra. Red lines: the spectra after smoothing with a 0.2 $\mu$Hz. Blue lines: Our best-fit MCMC model. Note that not all observed modes are expected to show clear splitting due to the stochastic nature of the oscillations.}
    \label{fig:figure5}
\end{figure*}

\subsection{Estimation of the Stellar Inclination} \label{subsec:Ed}
Figure~\ref{fig:figure5} shows the five strongest dipole modes ($l=1$) in the oscillation spectrum of Kepler-129. We modeled the oscillations using a superposition of Lorentzian functions:
\begin{equation}
\mathcal{P}(\nu)=\sum_{n}\sum_{m=-l}^{l} \frac{\varepsilon_{lm}(i_{s})H_{nlm}}{1+4[\frac{\nu-\nu_{nl0}-m\delta \nu_{s}}{ \Gamma_{nlm}}]^{2}}+\mathcal{B}(\nu)
\end{equation}
where $H_{nlm}$ and $\Gamma_{nlm}$ are the height and width of the Lorentzian profiles corresponding to every $m$ component ($m=-1,0,1$ when $l=1$). $\mathcal{B}(\nu)$ describes the background terms coming from granulations, stellar activities and photon noise. The inner sum runs over the $m$ components of each rotationally split multiplet, while the outer sum runs over all observed modes, in radial order $n$. Note that we consider five most significant dipole modes that ($n=15, 16, 17, 18, 19; l=1$). 

We use a Markov-Chain Monte-Carlo algorithm to fit all five observable modes simultaneously. The central frequency $\nu_{nl0}$ and mode height $H_{nlm}$ for are fitted for each mode, while the angle $\cos i_{s}$, linewidth $\Gamma_{nlm}$, rotational splitting $\delta v$ and noise floor $\mathcal{B}(\nu)$ are assumed to be the same for all five modes. Hence, the fitting includes a total of 14 parameters. We adopt Jeffreys prior for the mode heights and uniform priors for all other parameters with boundaries of $0<\cos i_{s}<1$, $0<\delta v < 10$, $0<\Gamma_{nlm}<10$ and $H_{nlm}>0$. Note that we uniformly sample in $\cos i_{s}$, which corresponds to an isotropic spin distribution. 

We use 50 walkers and performed $ 10^{4}$ iterations with each walker. The first $10\%$ of each chain is discarded for burn-in. 
\begin{figure}
    \centering
    \includegraphics[width=\linewidth]{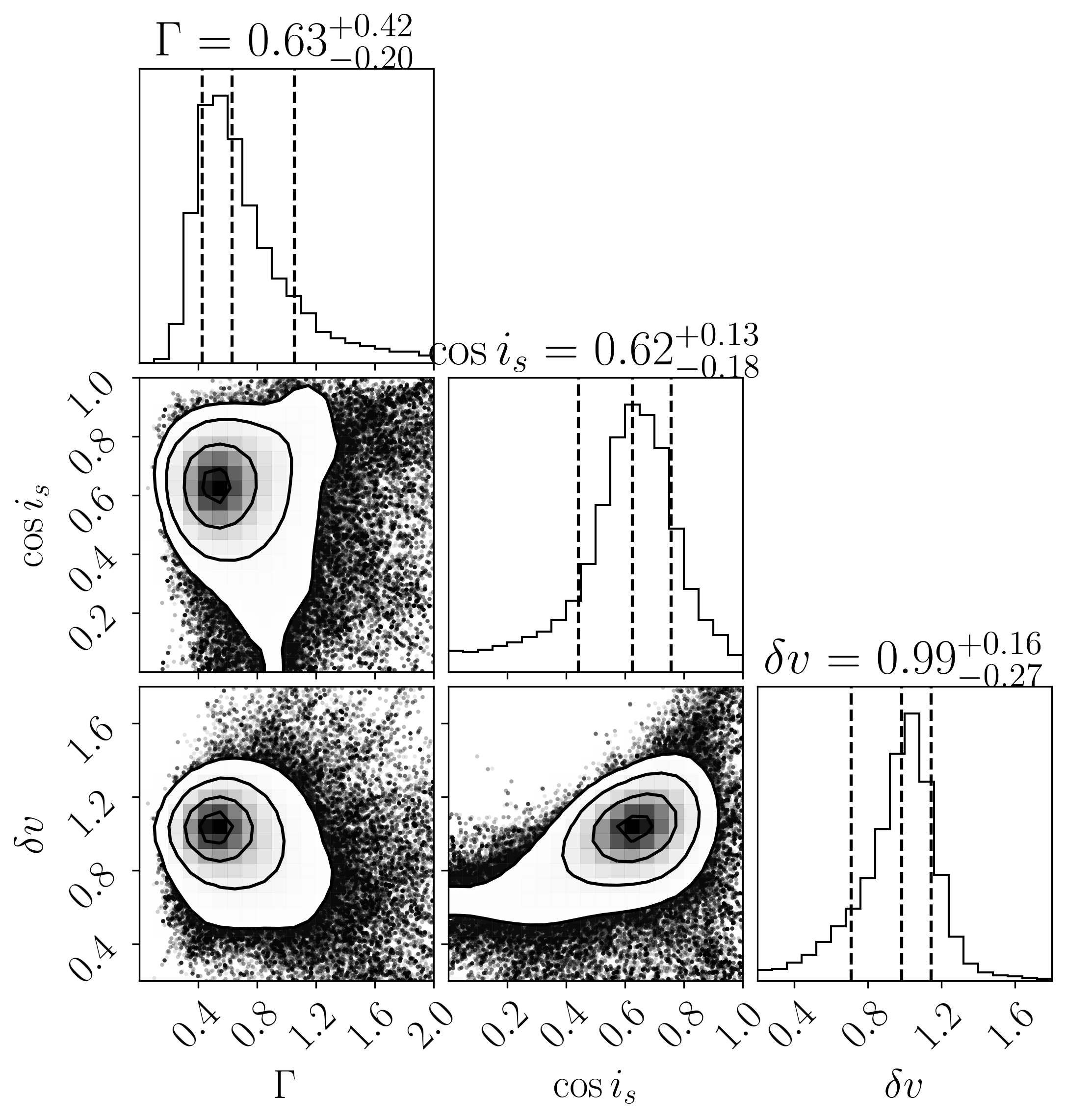}
    \caption{Joint posterior distributions for the $\Gamma$, $i_{s}$ and $\delta \nu$ from MCMC fitting. Moving forward, the solid lines correspond to 1$\sigma$,2$\sigma$ and 3$\sigma$ contours. The MCMC fitting converges at $\Gamma \sim 0.63\ \rm{\mu Hz} $, $\cos i_{s}\sim 0.62$ ($i_{s}\sim 52^{\circ}$) and $\delta \nu \sim 0.99\ \rm{\mu Hz}$, indicating the misalignment between stellar spin axis and orbital axes of the transiting planets Kepler-129 b and c. Our results show the data disfavour the spin-orbit alignment ($i_{s}=90^{\circ}$).}
    \label{fig:corner_astero}
\end{figure}
Figure~\ref{fig:corner_astero} presents the joint posterior distribution of $\Gamma$, $\cos i_{s}$ and $\delta v$. The MCMC fitting shows the best-fitting values of $\cos i_{s}=0.62^{+0.13}_{-0.18}$ , $\Gamma=0.63^{+0.42}_{-0.20}\ \rm{\mu Hz}$, and $\delta \nu=0.99^{+0.16}_{-0.27}\ \rm{\mu Hz} $. Hence we obtained $i_{s}=52^{+10}_{-13}$ deg. The best-fitting values and uncertainties were calculated as the median and $1\sigma$ interval of the marginalized posterior distribution for each parameter. Our results show the data disfavour the spin-orbit alignment ($i_{s}=90^{\circ}$) with $2\sigma$ confidence. 

Our results are consistent with that found by \citet{Campante2016} ($i_{s}=50^{+36.6}_{-15.6}$ deg at $1\sigma$ confidence), but have smaller uncertainties. We obtain a larger ratio $\delta \nu/ \Gamma$ than that in \citet{Campante2016} ($\delta \nu/ \Gamma \sim 0.6$). The both measured linewidth $\Gamma$ are consistent with the expected range ($0.9^{+0.4}_{-0.4}\ \rm{\mu Hz}$) at the effective temperature of Kepler-129 given by \cite{Lund2017}. \cite{Kamiaka2018} also measure the frequency splitting of Kepler-129 using Kepler data collected during Q6-Q7. They obtained a consistent result of $i_{s}=42.9^{+26.6}_{-23.2}$ deg ($1\sigma$) and $\delta \nu/ \Gamma \sim 0.95$. Note that they categorize Kepler-129 as a star for which seismic inclinations are difficult to measure, and hence our value should be used with some caution. However, given that we include more data collected during Q17, and there are two independent studies with similar results, we are confident about our results. In addition, using the stellar radius from \cite{SA2015} and the rotation frequency $\delta v$ from our MCMC fitting, we estimated stellar rotation velocity $v$ of $7.2^{+1.2}_{-1.9}\ \rm{km\ s^{-1}}$, which is significantly larger than projected velocity $v\sin i$ of $2.13\pm{1} \rm{km\ s^{-1}}$ measured from stellar spectra \citep{Petigura_thesis}. This indicates that $\sin i$ is much smaller than one and is consistent with our measurements of $i_{s}< 90^{\circ}$. 

\subsection{Limits on the Spin-Orbit Angle} \label{subsec:TrueO}
Measurements of $i_{s}$, along with the angle between the planet's orbital axis and line of sight ($i_{0}$) and the sky-projected spin-orbit angle ($\lambda$), can be used to compute the true spin-orbit angle $\psi$ as \citep{FW2009}
\begin{equation}\label{epsi}
    \cos{\psi}=\sin{i_{s}}\cos{\lambda}\sin{i_{0}}+\cos{i_{s}}\cos{i_{0}}
\end{equation}

The angle $i_{0}$ of a transiting planet is approximately $90^{\circ}$ given its nearly edge-on orbit. In principle, The sky-projected spin-orbit angle $\lambda$ can be obtained with measurements of the Rossiter-McLaughlin effect, although this angles is difficult to constrain for small planets. Based on our measurement of $i_{s}\sim52^{\circ}$ and assuming $i_{0}=90^{\circ}$ for the transiting planets Kepler-129 b and c, the lower limit for the true spin-orbit angle $\psi$ can be approximated with equation~\ref{epsi} when $\lambda$ is unknown:
\begin{equation}
    \cos{\psi} < \sin{i_{s}}\sin{i_{0}}+\cos{i_{s}}\cos{i_{0}}
\end{equation}
Hence, the true spin-orbit angle between the star and transiting planets should be larger than  $38^{\circ}$, indicating a misalignment between the orbital planes of Kepler-129 b and c and the stellar equatorial plane.

\section{Orbital Dynamics}\label{subsec:sspin}

The existence of Kepler-129 d offers a natural explanation for the spin-orbit misalignment of the inner planets. If the orbit of Kepler-129 d is inclined with respect to those of the inner planets, it could have imposed a torque on the inners and excited them out of the equatorial plane of the host star. In addition to exciting the spin-orbit angle, an inclined outer giant planet could also excite mutual inclinations between inner planets, possibly preventing them from transiting together \citep{Becker2017, read_transit_2017, DongandPu2017}. The fact that we observe both Kepler-129 b and c to be transiting places an additional constraint on the inclination of Kepler-129 d. In other words, the inclination of d must be large enough that a spin-orbit angle of $\sim38\degr$ can be produced, while small enough that both planets b and c have a large probability of transiting together. In this section, we investigate the dynamic evolution of both the spin-orbit angle (\ref{subsubsec:TVP}) and the mutual inclination between inner planets (\ref{subsec:deltai}). 
\subsection{Analytical Model}\label{subsec:5_1}
\subsubsection{Spin-Orbit Angle Evolution}\label{subsubsec:TVP}

We apply the results of the `three-vector problem' \citep{BL06,BL09,BF14I,BF14} to explore how the spin-orbit angle evolves under the influence of an inclined Kepler-129 d. The three-vector problem was developed to model the secular evolution of three coupled angular motions. Here, the three vectors are the angular momentum of the star $\vec{L_{\star}}=L_{\star}\hat{l}_{\star}$, the total angular momentum of the inner planets $\vec{L_{in}}=L_{in}\hat{l}_{in}$, and that of the outer giant planet $\vec{L_{d}}=L_{d}\hat{l}_{d}$, where $\hat{l}_{\star},\hat{l}_{in},\hat{l}_{d}$ are unit vectors. In this subsection, we consider the total angular momentum of the inner planetary system $\vec{L}_{in}$ instead of the two individual planets
\begin{equation}
\vec{L}_{in}=\sum_{j=b,c}L_{j}\hat{l}_{j}=L_{in}\hat{l}_{in}
\end{equation}
where $\hat{l}_{j}$ is the unit vector normal to the orbit of planet $j$. The three vectors would precess around each other during the evolution. Following the convention in \cite{BF14}, we denote $\nu_{1}, \nu_{2}, \nu_{3}, \nu_{4}$ as the precession frequencies of $\hat{l}_{\star}$ around $\hat{l}_{in}$, of $\hat{l}_{in}$ around $\hat{l}_{\star}$, of $\hat{l}_{in}$ around $\hat{l}_{d}$, and of $\hat{l}_{d}$ around $\hat{l}_{in}$, respectively. They can be estimated using the stellar  parameters ($M_{\star}$, $R_{\star}$, $P_{tot}$, $k_{2}$, $C$) and planetary properties and orbital parameters ($m_{b}$, $m_{c}$, $m\sin i_{d}$, $a_{b}$, $a_{c}$, $a_{d}$, $e_{d}$) (see in \cite{BF14}).

We find the Kepler-129 system is consistent with the ``Pure Orbital Regime'' in \citet{BF14}, with $\nu_{1},\nu_{2},\nu_{4}\ll \nu_{3}$. For Kepler-129, we estimate $\nu_{1}\approx \ 1.5\times10^{-4} \ \rm{deg}\ \rm{kyr}^{-1}$, $\nu_{2}\approx \ 5\times10^{-5} \ \rm{deg}\ \rm{kyr}^{-1}$, $\nu_{3}\approx \ 6.9 \ \rm{deg}\ \rm{kyr}^{-1}$, $\nu_{4}\approx \ 0.04 \ \rm{deg}\ \rm{kyr}^{-1}$. In this regime, the frequencies associated with the inner planetary system coupling to the stellar spin ($\nu_{1}$ and $\nu_{2}$) are much smaller than those associated with the inner planetary system coupling to the outer planet ($\nu_{3}$ and $\nu_{4}$). This suggests that the stellar spin would neither significantly influence the orbits of planets nor be affected by the motion of planets. Among the planets, Kepler-129 d contains much more angular momentum than Kepler-129 b and c, so its orbital plane is almost invariant ($\nu_{4}\ \ll\ \nu_{3} $). Therefore, the dominant evolution in Kepler-129 is the precession of inner planets around the orbital axis of the outer giant planet at a roughly constant angle. The precession period is $P=2\pi/\nu_{3}\approx 52$ kyr, which is much shorter than the stellar age.

\begin{figure}
    \centering
    \includegraphics[width=\linewidth]{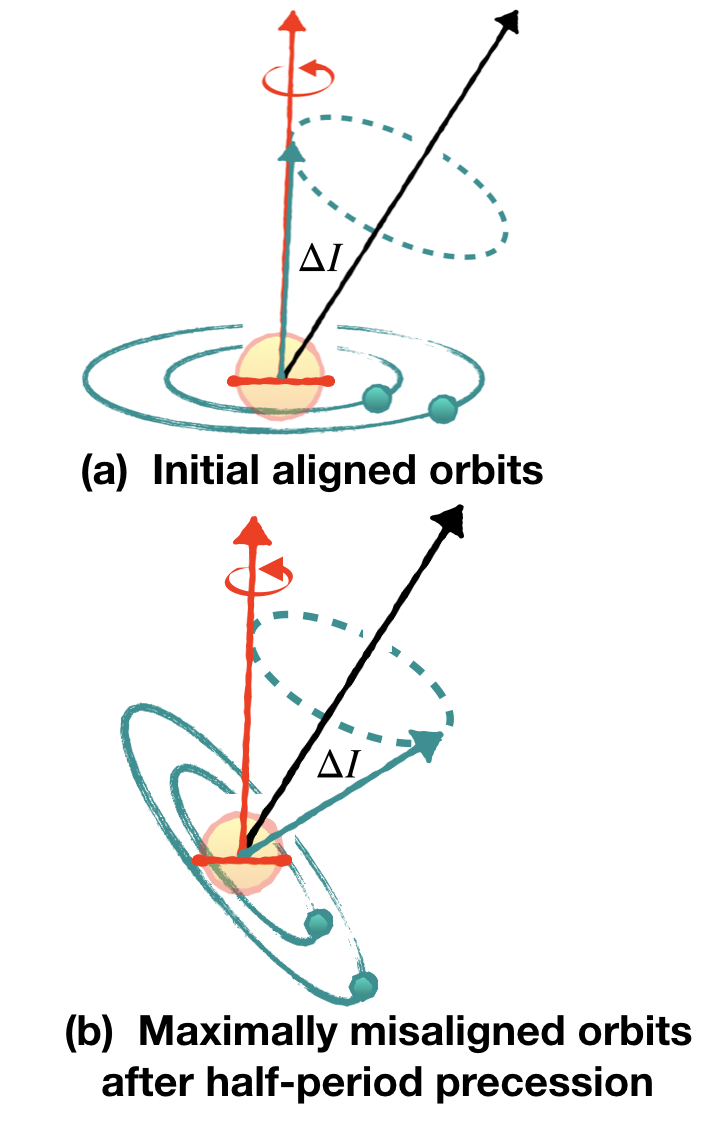}
    \caption{The spin-orbit angle of Kepler-129 b and c oscillates as their total angular momentum (green vector) precess around that of Kepler-129 d (black vector) at an angle $\Delta I$. $\Delta I$ keeps constant during the precession. (a) Initially, the inner planetary system is aligned with the stellar spin axis (red vector), but the outer giant planet is inclined by $\Delta I $ relative to them. (b) As the inner planets precess around outer giant planet (along the green dashed line), they achieve a maximum spin-orbit angle that is 2$\Delta I$.}

    \label{fig:figure91}
\end{figure}

We assume inner planets were aligned with the stellar spin axis when they formed as shown in Figure~\ref{fig:figure91} (a). The orbit of Kepler-129 d was inclined by $\Delta I$ relative to inner planets possibly due to the warped propoplanetary disk or through dynamical events such as planet-planet scattering. Then Kepler-129 b and c began to precess around Kepler-129 d at a constant angle $\Delta I$. During the inner planets' precession, their spin-orbit angle oscillates over time and reaches the maximum of $2\Delta I$ after half-period's precession (Figure~\ref{fig:figure91} (b)) \citep{BF14}. Therefore, our measurement of spin orbit angle ($\sim 38^{\circ}$) in Section~\ref{sec:AA} requires that Kepler-129 d is inclined relative to Kepler-129 b and c by at least $\sim19^{\circ}$.

\subsubsection{Mutual Inclination between Inner Planets} \label{subsec:deltai}

In this section, we investigate the efficacy with which Kepler-129 d excites mutual inclinations between planet b and c. Here, we neglect the influence of the stellar spin angular momentum as the precession rate of $\vec{L_{in}}$ around $\vec{L_{\star}}$ ($\nu_{2}$) is much slower than that of $\vec{L_{in}}$ around $\vec{L_{d}}$ ($\nu_{3}$) as found in Section~\ref{subsubsec:TVP}. In contrast to the above analysis, we now consider the orbital angular momenta of planet b and c separately, and thus have three vectors $\vec{L_{b}}=L_{b}\hat{l_{b}}$, $\vec{L_{c}}=L_{c}\hat{l_{c}}$ and $\vec{L_{d}}=L_{d}\hat{l_{d}}$, where $\hat{l_{b}}$, $\hat{l_{c}}$,and $\hat{l_{d}}$ are unit vectors. Because the angular momentum of Kepler-129 d is much larger than that of Kepler-129 b and c ($L_{b}, L_{c}\ll L_{d}$), $\hat{l_{d}}$ is hardly affected by $L_{b}$ and $L_{c}$ and is approximately fixed. The evolution of $\hat{l_{b}}$ and $\hat{l_{c}}$ can be described as their precession around each other and around $\hat{l_{d}}$ \citep{DongandPu2017} 
\begin{equation}
    \frac{d\hat{l_{b}}}{dt}=-\nu_{bc}(\hat{l_{b}}\cdot \hat{l_{c}})(\hat{l_{c}}\times \hat{l_{b}})-\nu_{bd}(\hat{l_{b}}\cdot \hat{l_{d}})(\hat{l_{d}}\times \hat{l_{b}})
\end{equation}
\begin{equation}
    \frac{d\hat{l_{c}}}{dt}=-\nu_{cb}(\hat{l_{c}}\cdot \hat{l_{b}})(\hat{l_{b}}\times \hat{l_{c}})-\nu_{cd}(\hat{l_{c}}\cdot \hat{l_{d}})(\hat{l_{d}}\times \hat{l_{c}})
\end{equation}
where $\nu_{bc}$ and $\nu_{bd}$ represent the precession rate of $\hat{l_{b}}$ around $\hat{l_{c}}$ (driven by Kepler-129 c) and that of $\hat{l_{b}}$ around $\hat{l_{d}}$ (driven by Kepler-129 d). $\nu_{cb}$ and $\nu_{cd}$ are the precession rates of $\hat{l_{c}}$ around $\hat{l_{b}}$ (driven by Kepler-129 b) and around $\hat{l_{d}}$ (driven by Kepler-129 d). They can be computed using the masses, semi-major axes and angular momenta of the planets (See \cite{DongandPu2017}).

Together, these four parameters determine whether the two inner planets dynamically couple with each other. A difference between $\nu_{bd}$ and $\nu_{cd}$ means that two inner planets precess around the giant planet at different rates, resulting in the separation of $\hat{l_{b}}$ from $\hat{l_{c}}$. On the other hand, the precession of Kepler-129 b and c around each other (at rates of $\nu_{bc}$ and $\nu_{cb}$) act to keep $\hat{l_{b}}$ and $\hat{l_{c}}$ coupled together. \cite{DongandPu2017} define the parameter $\epsilon=(\nu_{bd}-\nu_{cd})/ (\nu_{bc}+\nu_{cb})$ to estimate the relative coupling strength between the inner planets compared to the `disruptive' force of the outer planet. If $\epsilon \gg 1 \ (\nu_{cd}-\nu_{bd} \gg \nu_{bc}+\nu_{cb} )$, $\hat{l_{b}}$ and $\hat{l_{c}}$ will be forced apart by their different precession rates around planet d and thereby acquire relatively large mutual inclinations. In this case, their maximum mutual inclination will be two times of the inclination of outer planet $i_{d}$ ($i_{bc, max}=2 i_{d}$) \citep{DongandPu2017}. Conversely, if $\epsilon \ll 1 \ (\nu_{cd}-\nu_{bd} \ll \nu_{bc}+\nu_{cb} )$, the two inner planets will strongly couple and precess around Kepler-129 d together. In the condition of $\epsilon \ll 1$, their maximum mutual inclination can be given by the product of $\epsilon$ and $\sin(2i_{d})$ ($\mid \sin(i_{bc, max})\mid=\epsilon \mid \sin 2i_{d} \mid$) \citep{DongandPu2017}.

For the Kepler-129 planets, we find $\nu_{bc}\approx \ 24 \ \rm{deg}\ \rm{kyr}^{-1}$, $\nu_{bd}\approx \ 1.4 \ \rm{deg}\ \rm{kyr}^{-1}$, $\nu_{cb}\approx \ 2.5 \ \rm{deg}\ \rm{kyr}^{-1}$, $\nu_{cd}\approx \ 7.4 \ \rm{deg}\ \rm{kyr}^{-1}$. As $\nu_{bc}+\nu_{cb}$ is nearly 4.5 times larger than $\nu_{cd}-\nu_{bd}$, this indicates that Kepler-129 b and c precess around Kepler-129 d together while keeping a relatively small mutual inclination. Here, we obtained a rough estimate of $i_{bc,max}\sim 8^{\circ}$ if the outer giant planet is inclined by $19^{\circ}$ relative to inner planets. 

\subsection{N body Simulation} \label{subsec:IE}

\subsubsection{Simulation Setup}\label{subsubsec:NSS}
We present the analytical model that an inclined outer giant planet can cause the spin-orbit misalignment of two inner planets in section~\ref{subsubsec:TVP}. But the mutual inclination between Kepler-129 d and inner planets cannot be determined with only RV observations due to the degeneracy between the true mass and inclination of Kepler-129 d. In this section, we performed N-body simulations to set constraints on the inclination between the inner planets and the outer giant planet.  

The numerical integrations are carried out using the N-body package \textit{REBOUND} \citep{Rein2012}. We use the stellar mass and radius from \cite{SA2015} for the simulation setup. The initial conditions of planets ( $m_{c}$, $m\sin i_{d}$, $a_{d}$, $e_{d}$, $\omega_{d}$) are drawn from the MCMC posterior samples obtained in Section~\ref{sec:Kfit} (see Table~\ref{tab:star-compare}). For Kepler-129 b, we drew $m_{b}$ from a Gaussian distribution that is centered at the predicted mass given by mass radius relation in \cite{WM2014} and $3 \sigma$ below the measured upper limit ($\mu=1.38\ M_{\oplus},\sigma=0.4\ M_{\oplus}$). We set $a_{b}=0.13$ AU, $a_{c}=0.39$ AU based on transit observations \citep{Van2015}. We also set their eccentricities and argument of pericenter ($e_{b}$, $e_{c}$, $\omega_{b}$, $\omega_{c}$) to 0 since transit observations show both planet b and c are consistent with having zero eccentricity \citep{Van2015}. Assuming Kepler-129 b and c are nearly coplanar in the beginning, we drew the initial mutual inclination between Kepler-129 b and c from a Rayleigh distribution with a width of $1.5^{\circ}$ \citep{Fabrycky2014} and set their initial longitude of ascending node $\Omega$ to be zero. We did not use the inclinations from transit observations because the published impact parameters tend to be poorly constrained and highly degenerate with eccentricities and limb darkening models. Furthermore, it is not essential to begin our simulations with the exact values of the inclinations that the planets currently have because the mutual inclinations of the planets evolve with time. The simulations are divided into six samples, which have initial mutual inclinations between the outermost inner planet and the giant $i_{cd,0}$ as $5^{\circ}$, $10^{\circ}$, $15^{\circ}$,  $20^{\circ}$, $25^{\circ}$, $30^{\circ}$. We repeated 1000 trials for each sample, which amounts to 6000 simulations in total, and run every simulation for 200 kyr.

\subsubsection{Inclination Oscillation of Inner Planets}\label{subsubsec:inc}

\begin{figure*}
    \centering
    \includegraphics[width=0.9\linewidth]{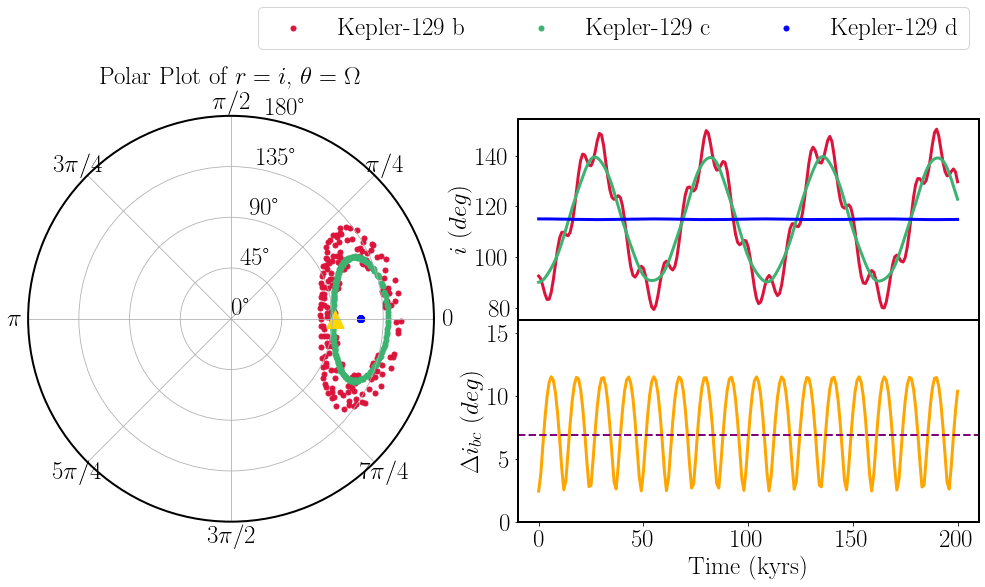}
    \caption{An example of the dynamical evolution in Kepler-129 from our N-body simulations. The simulation setup is: $m_{b}=3.4\ \rm{M_{\earth}}, m_{c}=47\ \rm{M_{\earth}}, m_{d}=8.3\ \rm{M_{Jup}}, a_{d}=4\ \rm{AU}, i_{cd,0}=25^{\circ}, i_{bc,0}=2.4^{\circ}$. The initial mutual inclination between planet b and c is drawn from a Rayleigh distribution with a width of $1^{\circ}.5$. Left: Polar plot of r = $i$ (inclination), $\theta = \Omega$ (longitude of ascending node). The orbit of Kepler-129 d is nearly invariant (blue), whereas Kepler-129 b (red) and c (green) both precess around Kepler-129 d. As Kepler-129 b and c precess, their inclinations oscillate between the initial value ($\sim 90^{\circ}$, marked by the yellow triangle) and $\sim 150^{\circ}$. In the meantime, Kepler-129 b precesses around Kepler-129 c at a much faster rate, which shows up as short-period variations in the red trajectory. Top right: Inclination of the three planets as a function of time. Bottom right: Mutual inclination between planet b and c $\Delta i_{bc}$ change as a function of time. Kepler-129 b and c both remain in a transiting configuration if $\Delta i_{bc}$ is below the threshold ($6.9^{\circ}$) marked by the purple dashed line.}
    \label{fig:figure93}
\end{figure*}

We show the orbital evolution of three planets in the left panel of Figure~\ref{fig:figure93}, which is a polar plot where the radical coordinate is inclination ($r=i$), and the angular coordinate is longitude of ascending node ($\theta=\Omega$). The overall behaviour matches the predictions of the analytical model in \ref{subsec:5_1}. First, the location of Kepler-129 d (blue points) stays nearly constant, indicating its orbit is approximately invariant during the entire simulation. Second, the orbits of Kepler-129 b (red) and c (green) trace cyclical trajectories around Kepler-129 d, suggesting they precess around the orbital axis of Kepler-129 d together. These are consistent with the analytical argument that Kepler-129 d contains much more angular momentum than Kepler-129 b and c ($\nu_{4}\ \ll\ \nu_{3} $), so the orbit of Kepler-129 d is hardly influenced while Kepler-129 b and c precess around Kepler-129 d together. During the precession, the inclination of Kepler-129 b and c oscillate as a function of time with a period of $\sim 54$ kyr (also see top right panel of Figure~\ref{fig:figure93}). This time scale is close to the analytic estimate in \ref{subsubsec:TVP} ($2\pi/\nu_{3}\approx $ 52 kyr). In addition, Kepler-129 b also precesses around Kepler-129 c at a faster rate, which appears as shorter-period variations in the red trajectory. The behavior is predicted when we consider angular momenta of Kepler-129 b and c separately in \ref{subsec:deltai}. This short period of $\sim 12$ kyr is also consistent with the precession period of Kepler-129 b around Kepler-129 c estimated in \ref{subsec:deltai} ($2\pi/\nu_{bc}\approx 15$ kyr).

\begin{figure}
    \centering
    \includegraphics[width=\linewidth]{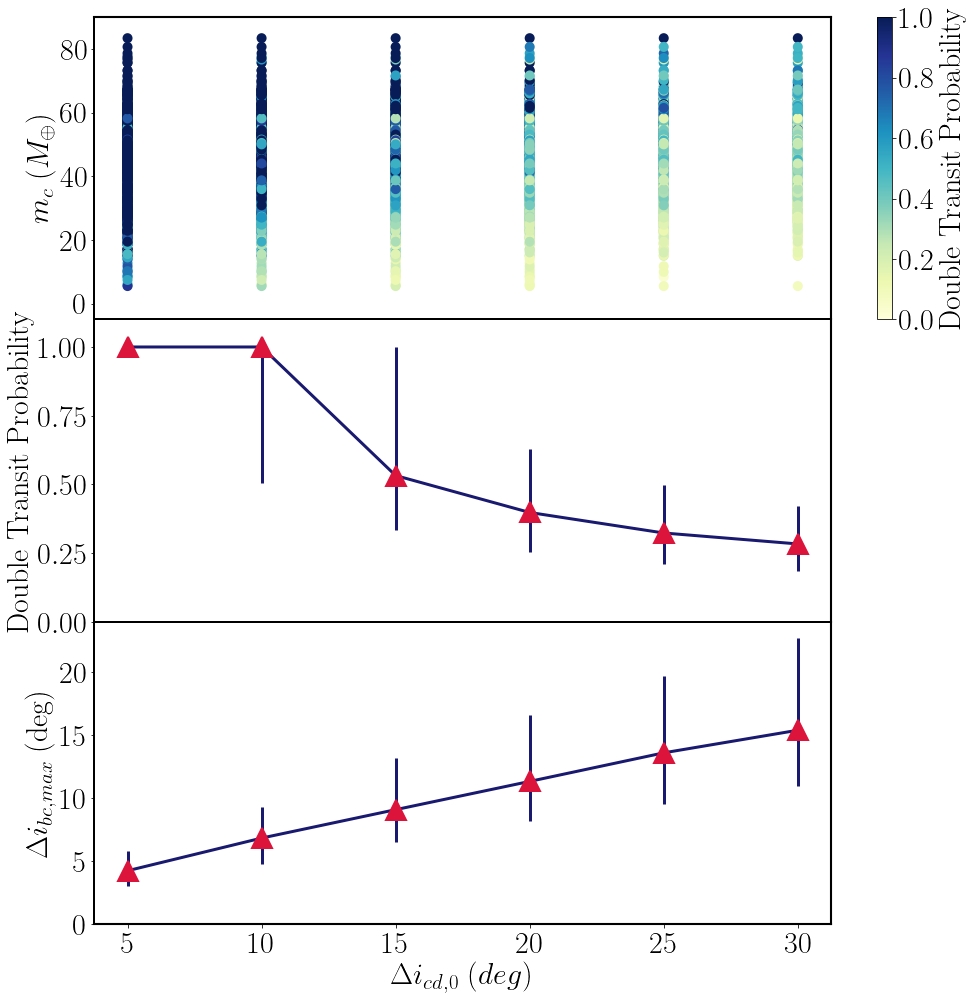}
    \caption{Mutual inclination between Kepler-129 b and c increases, and thus the double transit probability of them decreases as the misalignment between inner and outer planets rises. Top: Double transit probability of simulations with initial mutual inclination between Kepler-129 c and d ($\Delta i_{cd,0}$) ranging from $5^{\circ}$ to $30^{\circ}$. The mass of Kepler-129 c $m_{c}$ is drawn from the MCMC posterior distribution obtained in \ref{subsec:br}.  Middle and Bottom: Median value of the double transit probability and $\Delta i_{bc,max}$ of samples with the same $\Delta i_{cd,0}$ in the top panel, with error bars corresponding to the $1\sigma$ confidence interval. When $i_{cd,0}=5^{\circ}$, $\Delta i_{bc,max}$ is around $4^{\circ}$, rising linearly to $i_{bc,max}=15^{\circ} $ when $\Delta i_{cd,0}=30^{\circ}$. Accordingly, the double transiting probability decreases as $\Delta i_{cd,0}$ increases. Here, the inner planets will be observed as both transiting $\sim 40\%$ of time if $\Delta i_{cd,0}=20^{\circ}$ and less than $\sim 28\%$ of time if $\Delta i_{cd,0}=30^{\circ}$. In addition, at a given $\Delta i_{cd,0}$, the double transiting probability increases as $m_{c}$ increases, indicating that a more massive Kepler-129 c enables stronger coupling between the inner planets which helps to resist perturbations from Kepler-129 d.}
    \label{fig:figure8}
\end{figure}
\subsubsection{Double Transit Probability} \label{subsubsec:DTP}

In order to quantify whether the simulated inner planets transit together as observed, we define the parameter Double Transit Probability (DTP) for Kepler-129 b and c \citep[e.g.][]{read_transit_2017,Becker2017}. DTP is the fraction of time when two inner planets can both be observed to transit from any line of sight along their mutual ecliptic. This occurs when their mutual inclination is lower than a threshold $\Delta i_{max}= \arctan(R_{\star}/a_{b})+ \arctan(R_{\star}/a_{c})$. The two planets attain $\Delta i_{max}$ in the extreme scenario where they present grazing transits in two different hemispheres of the host star. Note that this definition does not assume a specific line of sight, e.g. from Earth.  We still consider planets to be both transiting if they can been seen from another line of sight, but not from Earth's line of sight. In our case, $\Delta i_{max}=6.9^{\circ}$ for Kepler-129 b and c, as marked by the dashed line in the bottom right panel of Figure~\ref{fig:figure93}. For each of our simulations, we calculated DTP as the fraction of time when $\Delta i_{bc}<\Delta i_{max}=6.9^{\circ}$, where $\Delta i_{bc}$ is the mutual inclination between planet b and c. 

DTP depends on the architecture of the system. For example, \cite{Becker2017} found that exterior perturbers with smaller periastron distances are more likely to disturb compact inner planets and excited them out of a mutually transiting configuration. In our case, we constrain the orbital elements of three planets from observations in Section~\ref{subsec:br} and \cite{Van2015} (see Table~\ref{tab:star-compare}). One missing but important element of the system is the mutual inclination between planets.  Hence we vary the initial mutual inclination between the outermost transiting planet Kepler-129 c and the giant planet Kepler-129 d ($i_{cd,0}$) from $5^{\circ}$ to $30^{\circ}$  to explore its effect on DTP. In turn, we get a constrain on the $i_{cd,0}$ based on DTP. Here, $i_{cd,0}$ is approximately the initial inclination between the inner planetary system and Kepler-129 d since planets b and c are almost co-planar in the beginning of the simulation. Note that we limit $i_{cd,0}$ to below $30^{\circ}$ to avoid Kozai-Lidov oscillations that can cause the inclination and eccentricity of inner planets to attain high values, and may even de-stabilize their orbits. 

In the top panel of Figure~\ref{fig:figure8}, we plot DTP against $i_{cd,0}$ and the mass of Kepler-129 c $m_{c}$.  In the middle and bottom panel, we plot the median value and $1 \sigma$ confidence interval of DTP and $i_{bc,max}$ of samples with the same $i_{cd,0}$ in the top panel. The dispersion in DTP is largely from the different values of $m_{c}$, which were drawn from posteriors of the RV fit. Overall, DTP decreases with increasing $i_{cd,0}$ because a more inclined perturber imposes a larger torque on the inner planets and excites larger mutual inclinations between them. When $i_{cd,0}<10^{\circ}$, DTP is equal to or around 1, implying that the co-planarity of Kepler-129 b and c is not disturbed if the outer giant planet is inclined by less than $10^{\circ}$ relative to them. In Section~\ref{subsubsec:TVP}, we found that the spin-orbit angle $ > \sim 38^{\circ}$  between the inner planets and the stellar spin requires at least $\sim 19^{\circ}$ of misalignment between the inner planetary system and outer giant planet. Our results shows that the probability of observing the simultaneous transits of Kepler-129 b and c is $\sim 40\%$ when $\Delta i_{cd,0}=20^{\circ}$. If $\Delta i_{cd,0}=30^{\circ}$, Kepler-129 b and c will remain in double transit configuration less than 28\% of the time, making them less likely to be observed as co-transiting.

In the analysis of Section~\ref{subsec:deltai}, we concluded that the orbital evolution of Kepler-129 b and c are determined by two effects: the separating `force' due to perturbations from Kepler-129 d and the coupling `force' due to their mutual precession around each other. In our simulations, we found a consistent result: DTP increases as the mass of Kepler-129 c ($m_{c}$) increases. Specifically, a more massive Kepler-129 c imposes a larger gravitational influence on Kepler-129 b and makes Kepler-129 b precess around planet c's angular momentum vector faster. In this way, the dynamical coupling between Kepler-129 b and c is stronger, making it more difficult for the giant planet to excite mutual inclinations between the inner planets.

In conclusion, we found that the mutual inclination between inner planets and the outer giant planet should be larger than $19^{^\circ}$ to produce the measured $\sim 38^{\circ}$ spin-orbit angle. But when the mutual inclination between inner planets and the outer giant planet is larger than $~30^{^\circ}$, the probability that Kepler-129 b and c both remain in transit configuration is smaller than 28\%. 
\section{Discussion}\label{DS}

\subsection{Comparison to Other Planetary Systems}
Several exoplanets systems that host spin-orbit misaligned close-in small planets have been found to contain an distant giant planet e.g. HAT-P-11 \citep{Winn2010,Yee2018}, $\pi$ Men \citep{Jones2002,Kunovac2020},  Kepler-56 \citep{Huber2013,Otor2016}, and WASP-107 \citep{Rubenzahl_2021,Piaulet2021}. Kepler-129 joins their ranks. HAT-P-11, $\pi$ Men and WASP-107 host single transiting planets, whereas Kepler-56 and Kepler-129 host two dynamically coupled transiting planets. In both HAT-P-11 and $\pi$ Men, the outer giant planets are highly inclined relative to the inner planets ($\sim 50^{\circ}$) and have high eccentricities ($\sim 0.6$) \citep{Xuan2020}. On the contrary, the giant planets in Kepler-56 and Kepler-129 have moderate eccentricities ($0.21$ and $0.15$, respectively). This current sample is consistent with the scenario that more eccentric and inclined perturbers could possibly excite some of the inner planets out of a transiting configuration, resulting in lower transit multiplicities. Furthermore, the outer transiting planets in Kepler-56 and Kepler-129 both have relatively high masses (Kepler-56 c $\sim195\ M_{\earth}$ and Kepler-129 c $\sim43\ M_{\earth}$), which enables strong coupling between themselves and the inner transiting planets. This effect suppresses the excitation of mutual inclinations between the two transiting planets from the outer giant.

\subsection{Possible Mechanisms to Cause the Spin-Orbit Misalignment}
We have discussed how an inclined giant planet could perturb the orbits of inner planets and cause their orbits to be misaligned with the stellar spin axis. But what gives the mutual inclination between the inner and outer planets in these systems? One possible scenario is that all planets form in the protoplanetary disc and are aligned with each other. At some point, two or more giant planets underwent dynamical encounters and only one giant planet remains, which ends up with a high eccentricity and high inclination orbit relative to the initial disk plane \citep{chatt2008}. The giant planets in HAT-P-11 and $\pi$ Men are consistent with this mechanisms with high eccentricities and high inclinations. An alternative possibility is that the planets formed in a warped protoplanetary disc with misaligned inner and outer components \citep{nealon2019, Xuan2020}. In this scenario, the inner and outer planets could be misaligned with each other from the beginning. The two scenarios are both possible for Kepler-129 system. Since the planet-planet scattering may produce relatively high eccentricities for the remaining giant planets whereas a warped protoplanetary disc does not, more discoveries of such systems in the future can help to study the eccentricities distribution of the outer giant planets statistically and distinguish the two possibilities.

Several other mechanisms can also explain the spin-orbit misalignment. One possibility is that the stellar spin was initially misaligned with respect to the protoplanetary disk. In this case, the inner planets may be misaligned with the stellar spin when they formed. \cite{Hjorth2021} found two co-planar transiting planets' orbits are retrograde with respect to the host star (K2-290A) in a triple system, indicating the protoplanetary disk was misaligned due to perturbations from the neighbouring star.  \cite{Spalding2014} and \cite{Spalding2020} argue that the quadrupolar gravitational potential of a tilted, rapidly rotating host star would torque the orbits of close-in planets. Because Kepler-129 has already evolved into the sub-giant stage and rotates relatively slowly, it is more likely that the influence of Kepler-129 vanishes as it spins down. 

\subsection{Opportunities for Future Observation}
Our RV baseline is too short to cover a full orbital period of Kepler-129 d. Future RV monitoring will provide a better constraint on the orbital period and minimum mass of Kepler-129 d. In addition, more RV data would enable more precise masses for Kepler-129 b and c. 
The Transiting Exoplanet Survey Satellite (\textit{TESS}) has the potential to reveal more transiting planets and its short cadence data can also be used to measure spin-orbit angles with asteroseismology for other planetary systems (TESS will not be able to detect oscillations for Kepler-129 since it is too faint). Combining results from \textit{TESS} and long-baseline RV observations could help us uncover other systems like Kepler-129 and study the influence of outer giant planets on small inner planets on a statistical level. 

\section{Conclusion} \label{sec:Ds}
We have presented the discovery of a long-period giant planet Kepler-129 d outside two known transiting sub-Neptune sized planets Kepler-129 b and c, and studied the orbital dynamics of the system. Our main conclusions are as follows:

\begin{itemize}
    \item Kepler-129 hosts two known transiting planets Kepler-129 b ($P_{b}=15.79$ days, $r_{b}=2.40\pm{0.04}\ R_{\earth}$) and Kepler-129 c ($P_{c}=82.20$ days, $r_{c}=2.52\pm{0.07}\ R_{\earth}$). We constrain the masses of Kepler-129 b and c with RV observations: $m <20\ M_{\earth}, m=43^{+13}_{-12}\ M_{\earth}$.
    \item Kepler-129 d is a long-period giant planet with moderate eccentricity ($P_{d}=7.2^{+0.4}_{-0.3}$ yr, $e_{d}=0.15^{+0.07}_{-0.05} $) outside the compact inner system. Kepler-129 d is a massive planet ($m\sin i_{d}=8.3^{+1.1}_{-0.7}\ M_{Jup}$), whose minimum mass is close to the traditional boundary between planets and brown dwarfs. The true mass of Kepler-129 d may be larger due to the unknown inclination so we can not rule out the possibility that it is a brown dwarf. 
    \item  Kepler-129 is a subgiant star with a clear presence of oscillation modes. From our best-fit models to the stellar oscillations, we found that the angle between stellar spin axis and line of sight to be $i_{s}=52^{+10}_{-13}$ deg. Assuming Kepler-129 b and c have edge-on orbits, the spin-orbit angle of inner planets is $>39^{\circ}$.
    \item The spin-orbit misalignment of Kepler-129 b and c indicates that their orbits may have been tilted via nodal precession around a misaligned Kepler-129 d. This scenario requires a mutual inclination between the inner planetary system and Kepler-129 d at least $19^{\circ}$.
    \item N-body simulations show Kepler-129 b and c both remain transiting 40\% of the time if they are inclined by $20^{\circ}$ relative to Kepler-129 d. This due to the relatively strong coupling between the two inner planets. However, if their inclination relative to Kepler-129 d rises to $30^{\circ}$ then Kepler-129 b and c will be observed to transit together only 28\% of the time. 
    
\end{itemize}

\acknowledgments

J.Z. would like to thank Daniel Fabrycky, Juliette  Becker, Dong Lai, Johanna Teske, and Jerry Xuan for helpful discussions. L.M.W. is supported by the Beatrice Watson Parrent Fellowship and NASA ADAP Grant 80NSSC19K0597.  D.H. acknowledges support from the Alfred P. Sloan Foundation, the National Aeronautics and Space Administration (80NSSC19K0597), and the National Science Foundation (AST-1717000). We also thank Tiago Campante for providing the best-fitting asteroseismic model for Kepler-129. M.R.K is supported by the NSF Graduate Research Fellowship, grant No. DGE 1339067. The authors wish to recognize and acknowledge the very significant cultural role and reverence that the summit of Mauna Kea has always had within the indigenous Hawaiian community. We are most fortunate to have the opportunity to conduct observations from this mountain.  This research has made use of the NASA Exoplanet Archive, which is operated by the California Institute of Technology, under contract with the National Aeronautics and Space Administration under the Exoplanet Exploration Program.

\vspace{5mm}
\facilities{Keck(HIRES), Kepler}


\software{\textit{Radvel} \citep{Fulton2018},
          \textit{Lightkurve} \citep{LK2018},  
          \textit{REBOUND} \citep{Rein2012} }
\clearpage
\newpage

\appendix

\begin{figure*}[!b]
    \centering
    \includegraphics[width=\linewidth]{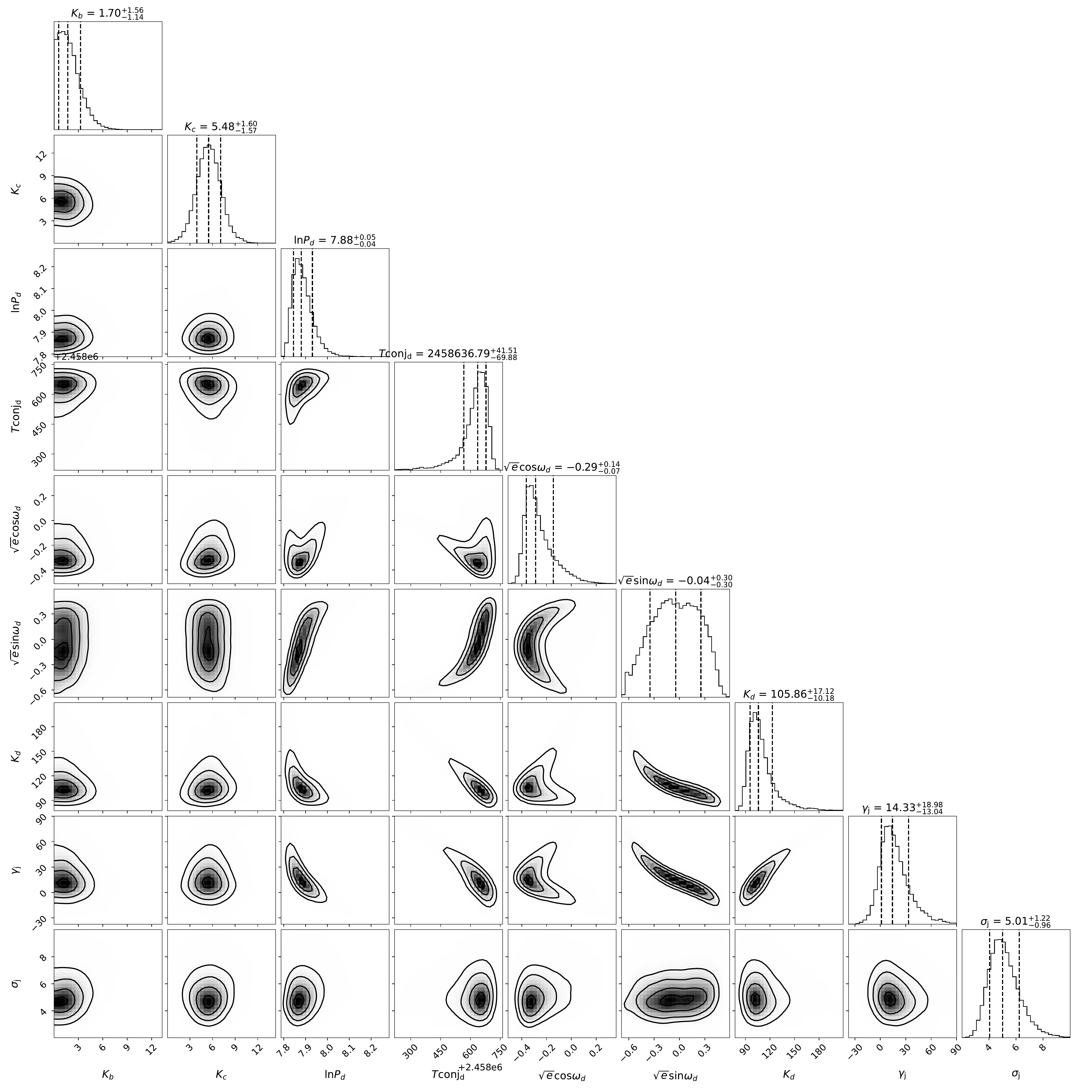}
    \caption{Joint posterior distributions for Kepler-129 b, c and d's orbital parameters ($P_{d}$, $K_{d}$, $e_{d}$, $\omega_{d}$, $tc_{d}$, $K_{b}$, $K_{c}$, $\gamma$, $\sigma$) using MCMC method. Moving forward, the solid lines correspond to 1$\sigma$,2$\sigma$ and 3$\sigma$ contours.}
    \label{fig:figure3}
\end{figure*}

\clearpage
\newpage
\bibliography{sample63}{}

\begin{thebibliography}{}
\expandafter\ifx\csname natexlab\endcsname\relax\def\natexlab#1{#1}\fi
\providecommand{\url}[1]{\href{#1}{#1}}
\providecommand{\dodoi}[1]{doi:~\href{http://doi.org/#1}{\nolinkurl{#1}}}
\providecommand{\doeprint}[1]{\href{http://ascl.net/#1}{\nolinkurl{http://ascl.net/#1}}}
\providecommand{\doarXiv}[1]{\href{https://arxiv.org/abs/#1}{\nolinkurl{https://arxiv.org/abs/#1}}}

\bibitem[{{Armstrong} {et~al.}(2020){Armstrong}, {Lopez}, {Adibekyan}, {Booth},
  {Bryant}, {Collins}, {Deleuil}, {Emsenhuber}, {Huang}, {King}, {Lillo-Box},
  {Lissauer}, {Matthews}, {Mousis}, {Nielsen}, {Osborn}, {Otegi}, {Santos},
  {Sousa}, {Stassun}, {Veras}, {Ziegler}, {Acton}, {Almenara}, {Anderson},
  {Barrado}, {Barros}, {Bayliss}, {Belardi}, {Bouchy}, {Brice{\~n}o}, {Brogi},
  {Brown}, {Burleigh}, {Casewell}, {Chaushev}, {Ciardi}, {Collins},
  {Col{\'o}n}, {Cooke}, {Crossfield}, {D{\'\i}az}, {Mena}, {Demangeon}, {Dorn},
  {Dumusque}, {Eigm{\"u}ller}, {Fausnaugh}, {Figueira}, {Gan}, {Gand hi},
  {Gill}, {Gonzales}, {Goad}, {G{\"u}nther}, {Helled}, {Hojjatpanah}, {Howell},
  {Jackman}, {Jenkins}, {Jenkins}, {Jensen}, {Kennedy}, {Latham}, {Law},
  {Lendl}, {Lozovsky}, {Mann}, {Moyano}, {McCormac}, {Meru}, {Mordasini},
  {Osborn}, {Pollacco}, {Queloz}, {Raynard}, {Ricker}, {Rowden}, {Santerne},
  {Schlieder}, {Seager}, {Sha}, {Tan}, {Tilbrook}, {Ting}, {Udry},
  {Vanderspek}, {Watson}, {West}, {Wilson}, {Winn}, {Wheatley}, {Villasenor},
  {Vines}, \& {Zhan}}]{Armstrong}
{Armstrong}, D.~J., {Lopez}, T.~A., {Adibekyan}, V., {et~al.} 2020, \nat, 583,
  39, \dodoi{10.1038/s41586-020-2421-7}

\bibitem[{{Ballot} {et~al.}(2008){Ballot}, {Appourchaux}, {Toutain}, \&
  {Guittet}}]{Ballot2008}
{Ballot}, J., {Appourchaux}, T., {Toutain}, T., \& {Guittet}, M. 2008, \aap,
  486, 867, \dodoi{10.1051/0004-6361:20079343}

\bibitem[{{Ballot} {et~al.}(2006){Ballot}, {Garc{\'\i}a}, \&
  {Lambert}}]{Ballot2006}
{Ballot}, J., {Garc{\'\i}a}, R.~A., \& {Lambert}, P. 2006, \mnras, 369, 1281,
  \dodoi{10.1111/j.1365-2966.2006.10375.x}

\bibitem[{{Becker} \& {Adams}(2017)}]{Becker2017}
{Becker}, J.~C., \& {Adams}, F.~C. 2017, \mnras, 468, 549,
  \dodoi{10.1093/mnras/stx461}

\bibitem[{{Bou{\'e}} \& {Fabrycky}(2014{\natexlab{a}})}]{BF14I}
{Bou{\'e}}, G., \& {Fabrycky}, D.~C. 2014{\natexlab{a}}, \apj, 789, 110,
  \dodoi{10.1088/0004-637X/789/2/110}

\bibitem[{{Bou{\'e}} \& {Fabrycky}(2014{\natexlab{b}})}]{BF14}
---. 2014{\natexlab{b}}, \apj, 789, 111, \dodoi{10.1088/0004-637X/789/2/111}

\bibitem[{{Bou{\'e}} \& {Laskar}(2006)}]{BL06}
{Bou{\'e}}, G., \& {Laskar}, J. 2006, \icarus, 185, 312,
  \dodoi{10.1016/j.icarus.2006.07.019}

\bibitem[{{Bou{\'e}} \& {Laskar}(2009)}]{BL09}
---. 2009, \icarus, 201, 750, \dodoi{10.1016/j.icarus.2009.02.001}

\bibitem[{{Bourrier} {et~al.}(2018){Bourrier}, {Lovis}, {Beust}, {Ehrenreich},
  {Henry}, {Astudillo-Defru}, {Allart}, {Bonfils}, {S{\'e}gransan}, {Delfosse},
  {Cegla}, {Wyttenbach}, {Heng}, {Lavie}, \& {Pepe}}]{Bourrier2018}
{Bourrier}, V., {Lovis}, C., {Beust}, H., {et~al.} 2018, \nat, 553, 477,
  \dodoi{10.1038/nature24677}

\bibitem[{{Bowler} {et~al.}(2020){Bowler}, {Blunt}, \& {Nielsen}}]{Bowler2020}
{Bowler}, B.~P., {Blunt}, S.~C., \& {Nielsen}, E.~L. 2020, \aj, 159, 63,
  \dodoi{10.3847/1538-3881/ab5b11}

\bibitem[{{Campante} {et~al.}(2011){Campante}, {Handberg}, {Mathur},
  {Appourchaux}, {Bedding}, {Chaplin}, {Garc{\'\i}a}, {Mosser}, {Benomar},
  {Bonanno}, {Corsaro}, {Fletcher}, {Gaulme}, {Hekker}, {Karoff}, {R{\'e}gulo},
  {Salabert}, {Verner}, {White}, {Houdek}, {Brand {\~a}o}, {Creevey},
  {Do{\v{g}}an}, {Bazot}, {Christensen-Dalsgaard}, {Cunha}, {Elsworth},
  {Huber}, {Kjeldsen}, {Lundkvist}, {Molenda-{\.Z}akowicz}, {Monteiro},
  {Stello}, {Clarke}, {Girouard}, \& {Hall}}]{Campante2011}
{Campante}, T.~L., {Handberg}, R., {Mathur}, S., {et~al.} 2011, \aap, 534, A6,
  \dodoi{10.1051/0004-6361/201116620}

\bibitem[{{Campante} {et~al.}(2016){Campante}, {Lund}, {Kuszlewicz}, {Davies},
  {Chaplin}, {Albrecht}, {Winn}, {Bedding}, {Benomar}, {Bossini}, {Handberg},
  {Santos}, {Van Eylen}, {Basu}, {Christensen-Dalsgaard}, {Elsworth}, {Hekker},
  {Hirano}, {Huber}, {Karoff}, {Kjeldsen}, {Lundkvist}, {North}, {Silva
  Aguirre}, {Stello}, \& {White}}]{Campante2016}
{Campante}, T.~L., {Lund}, M.~N., {Kuszlewicz}, J.~S., {et~al.} 2016, \apj,
  819, 85, \dodoi{10.3847/0004-637X/819/1/85}

\bibitem[{{Chaplin} \& {Miglio}(2013)}]{Chaplin2013}
{Chaplin}, W.~J., \& {Miglio}, A. 2013, \araa, 51, 353,
  \dodoi{10.1146/annurev-astro-082812-140938}

\bibitem[{{Chaplin} {et~al.}(2013){Chaplin}, {Sanchis-Ojeda}, {Campante},
  {Handberg}, {Stello}, {Winn}, {Basu}, {Christensen-Dalsgaard}, {Davies},
  {Metcalfe}, {Buchhave}, {Fischer}, {Bedding}, {Cochran}, {Elsworth},
  {Gilliland}, {Hekker}, {Huber}, {Isaacson}, {Karoff}, {Kawaler}, {Kjeldsen},
  {Latham}, {Lund}, {Lundkvist}, {Marcy}, {Miglio}, {Barclay}, \&
  {Lissauer}}]{chaplin_etal2013}
{Chaplin}, W.~J., {Sanchis-Ojeda}, R., {Campante}, T.~L., {et~al.} 2013, \apj,
  766, 101, \dodoi{10.1088/0004-637X/766/2/101}

\bibitem[{{Chatterjee} {et~al.}(2008{\natexlab{a}}){Chatterjee}, {Ford},
  {Matsumura}, \& {Rasio}}]{chatterjee2008}
{Chatterjee}, S., {Ford}, E.~B., {Matsumura}, S., \& {Rasio}, F.~A.
  2008{\natexlab{a}}, \apj, 686, 580, \dodoi{10.1086/590227}

\bibitem[{{Chatterjee} {et~al.}(2008{\natexlab{b}}){Chatterjee}, {Ford},
  {Matsumura}, \& {Rasio}}]{chatt2008}
---. 2008{\natexlab{b}}, \apj, 686, 580, \dodoi{10.1086/590227}

\bibitem[{{Collier Cameron} {et~al.}(2010){Collier Cameron}, {Bruce}, {Miller},
  {Triaud}, \& {Queloz}}]{Cameron2010}
{Collier Cameron}, A., {Bruce}, V.~A., {Miller}, G.~R.~M., {Triaud},
  A.~H.~M.~J., \& {Queloz}, D. 2010, \mnras, 403, 151,
  \dodoi{10.1111/j.1365-2966.2009.16131.x}

\bibitem[{{Damasso} {et~al.}(2020){Damasso}, {Sozzetti}, {Lovis}, {Barros},
  {Sousa}, {Demangeon}, {Faria}, {Lillo-Box}, {Cristiani}, {Pepe}, {Rebolo},
  {Santos}, {Zapatero Osorio}, {Gonz{\'a}lez Hern{\'a}ndez}, {Amate},
  {Pasquini}, {Zerbi}, {Adibekyan}, {Abreu}, {Affolter}, {Alibert}, {Aliverti},
  {Allart}, {Allende Prieto}, {{\'A}lvarez}, {Alves}, {Avila}, {Baldini},
  {Bandy}, {Benz}, {Bianco}, {Borsa}, {Bossini}, {Bourrier}, {Bouchy}, {Broeg},
  {Cabral}, {Calderone}, {Cirami}, {Coelho}, {Conconi}, {Coretti}, {Cumani},
  {Cupani}, {D'Odorico}, {Deiries}, {Dekker}, {Delabre}, {Di Marcantonio},
  {Dumusque}, {Ehrenreich}, {Figueira}, {Fragoso}, {Genolet}, {Genoni},
  {G{\'e}nova Santos}, {Hughes}, {Iwert}, {Kerber}, {Knudstrup}, {Landoni},
  {Lavie}, {Lizon}, {Lo Curto}, {Maire}, {Martins}, {M{\'e}gevand}, {Mehner},
  {Micela}, {Modigliani}, {Molaro}, {Monteiro}, {Monteiro}, {Moschetti},
  {Mueller}, {Murphy}, {Nunes}, {Oggioni}, {Oliveira}, {Oshagh}, {Pall{\'e}},
  {Pariani}, {Poretti}, {Rasilla}, {Rebord{\~a}o}, {Redaelli}, {Riva}, {Santana
  Tschudi}, {Santin}, {Santos}, {S{\'e}gransan}, {Schmidt}, {Segovia},
  {Sosnowska}, {Span{\`o}}, {Su{\'a}rez Mascare{\~n}o}, {Tabernero}, {Tenegi},
  {Udry}, \& {Zanutta}}]{Damasso2020}
{Damasso}, M., {Sozzetti}, A., {Lovis}, C., {et~al.} 2020, arXiv e-prints,
  arXiv:2007.06410.
\newblock \doarXiv{2007.06410}

\bibitem[{{de Laplace}(1796)}]{laplace1796}
{de Laplace}, P.~S. 1796, {Exposition du syst{\`e}me du monde},
  \dodoi{10.3931/e-rara-497}

\bibitem[{{De Rosa} {et~al.}(2020){De Rosa}, {Dawson}, \&
  {Nielsen}}]{DeRosa2020}
{De Rosa}, R.~J., {Dawson}, R., \& {Nielsen}, E.~L. 2020, arXiv e-prints,
  arXiv:2007.08549.
\newblock \doarXiv{2007.08549}

\bibitem[{Denham {et~al.}(2019)Denham, Naoz, Hoang, Stephan, \&
  Farr}]{denham_hidden_2019}
Denham, P., Naoz, S., Hoang, B.-M., Stephan, A.~P., \& Farr, W.~M. 2019, MNRAS,
  482, 4146, \dodoi{10.1093/mnras/sty2830}

\bibitem[{{D{\'\i}az} {et~al.}(2020){D{\'\i}az}, {Jenkins}, {Feng}, {Butler},
  {Tuomi}, {Shectman}, {Thorngren}, {Soto}, {Vines}, {Teske}, {Dragomir},
  {Villanueva}, {Kane}, {Berdi{\~n}as}, {Crane}, {Wang}, \&
  {Arriagada}}]{D2020}
{D{\'\i}az}, M.~R., {Jenkins}, J.~S., {Feng}, F., {et~al.} 2020, \mnras, 496,
  4330, \dodoi{10.1093/mnras/staa1724}

\bibitem[{{Dziembowski}(1977)}]{Dz1977}
{Dziembowski}, W. 1977, \actaa, 27, 203

\bibitem[{{Fabrycky} \& {Winn}(2009)}]{FW2009}
{Fabrycky}, D.~C., \& {Winn}, J.~N. 2009, \apj, 696, 1230,
  \dodoi{10.1088/0004-637X/696/2/1230}

\bibitem[{{Fabrycky} {et~al.}(2014){Fabrycky}, {Lissauer}, {Ragozzine}, {Rowe},
  {Steffen}, {Agol}, {Barclay}, {Batalha}, {Borucki}, {Ciardi}, {Ford},
  {Gautier}, {Geary}, {Holman}, {Jenkins}, {Li}, {Morehead}, {Morris},
  {Shporer}, {Smith}, {Still}, \& {Van Cleve}}]{Fabrycky2014}
{Fabrycky}, D.~C., {Lissauer}, J.~J., {Ragozzine}, D., {et~al.} 2014, \apj,
  790, 146, \dodoi{10.1088/0004-637X/790/2/146}

\bibitem[{{Foreman-Mackey} {et~al.}(2013){Foreman-Mackey}, {Hogg}, {Lang}, \&
  {Goodman}}]{FM2013}
{Foreman-Mackey}, D., {Hogg}, D.~W., {Lang}, D., \& {Goodman}, J. 2013, \pasp,
  125, 306, \dodoi{10.1086/670067}

\bibitem[{{Fulton} \& {Petigura}(2018)}]{FP18}
{Fulton}, B.~J., \& {Petigura}, E.~A. 2018, \aj, 156, 264,
  \dodoi{10.3847/1538-3881/aae828}

\bibitem[{{Fulton} {et~al.}(2018){Fulton}, {Petigura}, {Blunt}, \&
  {Sinukoff}}]{Fulton2018}
{Fulton}, B.~J., {Petigura}, E.~A., {Blunt}, S., \& {Sinukoff}, E. 2018, \pasp,
  130, 044504, \dodoi{10.1088/1538-3873/aaaaa8}

\bibitem[{{Gizon} \& {Solanki}(2003)}]{Gizon2003}
{Gizon}, L., \& {Solanki}, S.~K. 2003, \apj, 589, 1009, \dodoi{10.1086/374715}

\bibitem[{{Grossman} \& {Graboske}(1973)}]{GandG1973}
{Grossman}, A.~S., \& {Graboske}, H.~C. 1973, \apj, 180, 195,
  \dodoi{10.1086/151954}

\bibitem[{{Hekker} {et~al.}(2010){Hekker}, {Barban}, {Baudin}, {De Ridder},
  {Kallinger}, {Morel}, {Chaplin}, \& {Elsworth}}]{Hekker2010}
{Hekker}, S., {Barban}, C., {Baudin}, F., {et~al.} 2010, \aap, 520, A60,
  \dodoi{10.1051/0004-6361/201014944}

\bibitem[{{Hirano} {et~al.}(2011){Hirano}, {Narita}, {Shporer}, {Sato}, {Aoki},
  \& {Tamura}}]{Hirano2011}
{Hirano}, T., {Narita}, N., {Shporer}, A., {et~al.} 2011, \pasj, 63, 531,
  \dodoi{10.1093/pasj/63.sp2.S531}

\bibitem[{{Hjorth} {et~al.}(2021){Hjorth}, {Albrecht}, {Hirano}, {Winn},
  {Dawson}, {Zanazzi}, {Knudstrup}, \& {Sato}}]{Hjorth2021}
{Hjorth}, M., {Albrecht}, S., {Hirano}, T., {et~al.} 2021, Proceedings of the
  National Academy of Science, 118, 2017418118, \dodoi{10.1073/pnas.2017418118}

\bibitem[{{Howard} {et~al.}(2010){Howard}, {Johnson}, {Marcy}, {Fischer},
  {Wright}, {Bernat}, {Henry}, {Peek}, {Isaacson}, {Apps}, {Endl}, {Cochran},
  {Valenti}, {Anderson}, \& {Piskunov}}]{Howard2010}
{Howard}, A.~W., {Johnson}, J.~A., {Marcy}, G.~W., {et~al.} 2010, \apj, 721,
  1467, \dodoi{10.1088/0004-637X/721/2/1467}

\bibitem[{{Huang} {et~al.}(2017){Huang}, {Petrovich}, \& {Deibert}}]{Huang2017}
{Huang}, C.~X., {Petrovich}, C., \& {Deibert}, E. 2017, \aj, 153, 210,
  \dodoi{10.3847/1538-3881/aa67fb}

\bibitem[{{Huber} {et~al.}(2013){Huber}, {Carter}, {Barbieri}, {Miglio},
  {Deck}, {Fabrycky}, {Montet}, {Buchhave}, {Chaplin}, {Hekker},
  {Montalb{\'a}n}, {Sanchis-Ojeda}, {Basu}, {Bedding}, {Campante},
  {Christensen-Dalsgaard}, {Elsworth}, {Stello}, {Arentoft}, {Ford}, {Gilliland
  }, {Handberg}, {Howard}, {Isaacson}, {Johnson}, {Karoff}, {Kawaler},
  {Kjeldsen}, {Latham}, {Lund}, {Lundkvist}, {Marcy}, {Metcalfe}, {Silva
  Aguirre}, \& {Winn}}]{Huber2013}
{Huber}, D., {Carter}, J.~A., {Barbieri}, M., {et~al.} 2013, Science, 342, 331,
  \dodoi{10.1126/science.1242066}

\bibitem[{{Jones} {et~al.}(2002){Jones}, {Paul Butler}, {Tinney}, {Marcy},
  {Penny}, {McCarthy}, {Carter}, \& {Pourbaix}}]{Jones2002}
{Jones}, H. R.~A., {Paul Butler}, R., {Tinney}, C.~G., {et~al.} 2002, \mnras,
  333, 871, \dodoi{10.1046/j.1365-8711.2002.05459.x}

\bibitem[{{Kamiaka} {et~al.}(2018){Kamiaka}, {Benomar}, \&
  {Suto}}]{Kamiaka2018}
{Kamiaka}, S., {Benomar}, O., \& {Suto}, Y. 2018, \mnras, 479, 391,
  \dodoi{10.1093/mnras/sty1358}

\bibitem[{{Kamiaka} {et~al.}(2019){Kamiaka}, {Benomar}, {Suto}, {Dai},
  {Masuda}, \& {Winn}}]{Kamiaka2019}
{Kamiaka}, S., {Benomar}, O., {Suto}, Y., {et~al.} 2019, \aj, 157, 137,
  \dodoi{10.3847/1538-3881/ab04a9}

\bibitem[{{Kant}(1755)}]{Kant1775}
{Kant}, I. 1755, {Allgemeine Naturgeschichte und Theorie des Himmels}

\bibitem[{{Kunovac Hod{\v{z}}i{\'c}} {et~al.}(2020){Kunovac Hod{\v{z}}i{\'c}},
  {Triaud}, {Cegla}, {Chaplin}, \& {Davies}}]{Kunovac2020}
{Kunovac Hod{\v{z}}i{\'c}}, V., {Triaud}, A. H.~M.~J., {Cegla}, H.~M.,
  {Chaplin}, W.~J., \& {Davies}, G.~R. 2020, arXiv e-prints, arXiv:2007.11564.
\newblock \doarXiv{2007.11564}

\bibitem[{{Lai} \& {Pu}(2017)}]{DongandPu2017}
{Lai}, D., \& {Pu}, B. 2017, \aj, 153, 42, \dodoi{10.3847/1538-3881/153/1/42}

\bibitem[{{Landin} {et~al.}(2009){Landin}, {Mendes}, \& {Vaz}}]{landin2009}
{Landin}, N.~R., {Mendes}, L.~T.~S., \& {Vaz}, L.~P.~R. 2009, \aap, 494, 209,
  \dodoi{10.1051/0004-6361:20078403}

\bibitem[{{Ledoux}(1951)}]{Ledoux1951}
{Ledoux}, P. 1951, \apj, 114, 373, \dodoi{10.1086/145477}

\bibitem[{{Lightkurve Collaboration} {et~al.}(2018){Lightkurve Collaboration},
  {Cardoso}, {Hedges}, {Gully-Santiago}, {Saunders}, {Cody}, {Barclay}, {Hall},
  {Sagear}, {Turtelboom}, {Zhang}, {Tzanidakis}, {Mighell}, {Coughlin}, {Bell},
  {Berta-Thompson}, {Williams}, {Dotson}, \& {Barentsen}}]{LK2018}
{Lightkurve Collaboration}, {Cardoso}, J.~V.~d.~M., {Hedges}, C., {et~al.}
  2018, {Lightkurve: Kepler and TESS time series analysis in Python},
  Astrophysics Source Code Library.
\newblock \doeprint{1812.013}

\bibitem[{{Lopez} \& {Fortney}(2014)}]{Lopez2014}
{Lopez}, E.~D., \& {Fortney}, J.~J. 2014, \apj, 792, 1,
  \dodoi{10.1088/0004-637X/792/1/1}

\bibitem[{{Lund} {et~al.}(2017){Lund}, {Silva Aguirre}, {Davies}, {Chaplin},
  {Christensen-Dalsgaard}, {Houdek}, {White}, {Bedding}, {Ball}, {Huber},
  {Antia}, {Lebreton}, {Latham}, {Handberg}, {Verma}, {Basu}, {Casagrande},
  {Justesen}, {Kjeldsen}, \& {Mosumgaard}}]{Lund2017}
{Lund}, M.~N., {Silva Aguirre}, V., {Davies}, G.~R., {et~al.} 2017, \apj, 835,
  172, \dodoi{10.3847/1538-4357/835/2/172}

\bibitem[{{Masuda} {et~al.}(2020){Masuda}, {Winn}, \& {Kawahara}}]{Masuda2020}
{Masuda}, K., {Winn}, J.~N., \& {Kawahara}, H. 2020, \aj, 159, 38,
  \dodoi{10.3847/1538-3881/ab5c1d}

\bibitem[{{McLaughlin}(1924)}]{M1924}
{McLaughlin}, D.~B. 1924, \apj, 60, 22, \dodoi{10.1086/142826}

\bibitem[{{Nealon} {et~al.}(2019){Nealon}, {Pinte}, {Alexander}, {Mentiplay},
  \& {Dipierro}}]{nealon2019}
{Nealon}, R., {Pinte}, C., {Alexander}, R., {Mentiplay}, D., \& {Dipierro}, G.
  2019, \mnras, 484, 4951, \dodoi{10.1093/mnras/stz346}

\bibitem[{{Otor} {et~al.}(2016){Otor}, {Montet}, {Johnson}, {Charbonneau},
  {Collier-Cameron}, {Howard}, {Isaacson}, {Latham}, {Lopez-Morales}, {Lovis},
  {Mayor}, {Micela}, {Molinari}, {Pepe}, {Piotto}, {Phillips}, {Queloz},
  {Rice}, {Sasselov}, {S{\'e}gransan}, {Sozzetti}, {Udry}, \&
  {Watson}}]{Otor2016}
{Otor}, O.~J., {Montet}, B.~T., {Johnson}, J.~A., {et~al.} 2016, \aj, 152, 165,
  \dodoi{10.3847/0004-6256/152/6/165}

\bibitem[{{Petigura}(2015)}]{Petigura_thesis}
{Petigura}, E.~A. 2015, PhD thesis, University of California, Berkeley

\bibitem[{{Piaulet} {et~al.}(2021){Piaulet}, {Benneke}, {Rubenzahl}, {Howard},
  {Lee}, {Thorngren}, {Angus}, {Peterson}, {Schlieder}, {Werner}, {Kreidberg},
  {Jaouni}, {Crossfield}, {Ciardi}, {Petigura}, {Livingston}, {Dressing},
  {Fulton}, {Beichman}, {Christiansen}, {Gorjian}, {Hardegree-Ullman}, {Krick},
  \& {Sinukoff}}]{Piaulet2021}
{Piaulet}, C., {Benneke}, B., {Rubenzahl}, R.~A., {et~al.} 2021, \aj, 161, 70,
  \dodoi{10.3847/1538-3881/abcd3c}

\bibitem[{{Pu} \& {Lai}(2018)}]{PL2018}
{Pu}, B., \& {Lai}, D. 2018, \mnras, 478, 197, \dodoi{10.1093/mnras/sty1098}

\bibitem[{Read {et~al.}(2017)Read, Wyatt, \& Triaud}]{read_transit_2017}
Read, M.~J., Wyatt, M.~C., \& Triaud, A. H. M.~J. 2017, Monthly Notices of the
  Royal Astronomical Society, 469, 171, \dodoi{10.1093/mnras/stx798}

\bibitem[{{Rein} \& {Liu}(2012)}]{Rein2012}
{Rein}, H., \& {Liu}, S.~F. 2012, \aap, 537, A128,
  \dodoi{10.1051/0004-6361/201118085}

\bibitem[{{Rossiter}(1924)}]{R1924}
{Rossiter}, R.~A. 1924, \apj, 60, 15, \dodoi{10.1086/142825}

\bibitem[{{Rowe} {et~al.}(2014){Rowe}, {Bryson}, {Marcy}, {Lissauer},
  {Jontof-Hutter}, {Mullally}, {Gilliland}, {Issacson}, {Ford}, {Howell},
  {Borucki}, {Haas}, {Huber}, {Steffen}, {Thompson}, {Quintana}, {Barclay},
  {Still}, {Fortney}, {Gautier}, {Hunter}, {Caldwell}, {Ciardi}, {Devore},
  {Cochran}, {Jenkins}, {Agol}, {Carter}, \& {Geary}}]{Rowe2014}
{Rowe}, J.~F., {Bryson}, S.~T., {Marcy}, G.~W., {et~al.} 2014, \apj, 784, 45,
  \dodoi{10.1088/0004-637X/784/1/45}

\bibitem[{Rubenzahl {et~al.}(2021)Rubenzahl, Dai, Howard, Chontos, Giacalone,
  Lubin, Rosenthal, Isaacson, Batalha, Crossfield, Dressing, Fulton, Huber,
  Kane, Petigura, Robertson, Roy, Weiss, Beard, Hill, Mayo, Mocnik, Murphy, \&
  Scarsdale}]{Rubenzahl_2021}
Rubenzahl, R.~A., Dai, F., Howard, A.~W., {et~al.} 2021, The Astronomical
  Journal, 161, 119, \dodoi{10.3847/1538-3881/abd177}

\bibitem[{{Scargle}(1982)}]{Scargle1982}
{Scargle}, J.~D. 1982, \apj, 263, 835, \dodoi{10.1086/160554}

\bibitem[{{Schlaufman}(2018)}]{Schlaufman2018}
{Schlaufman}, K.~C. 2018, \apj, 853, 37, \dodoi{10.3847/1538-4357/aa961c}

\bibitem[{{Silva Aguirre} {et~al.}(2015){Silva Aguirre}, {Davies}, {Basu},
  {Christensen-Dalsgaard}, {Creevey}, {Metcalfe}, {Bedding}, {Casagrande},
  {Handberg}, {Lund}, {Nissen}, {Chaplin}, {Huber}, {Serenelli}, {Stello}, {Van
  Eylen}, {Campante}, {Elsworth}, {Gilliland}, {Hekker}, {Karoff}, {Kawaler},
  {Kjeldsen}, \& {Lundkvist}}]{SA2015}
{Silva Aguirre}, V., {Davies}, G.~R., {Basu}, S., {et~al.} 2015, \mnras, 452,
  2127, \dodoi{10.1093/mnras/stv1388}

\bibitem[{{Spalding} \& {Batygin}(2014)}]{Spalding2014}
{Spalding}, C., \& {Batygin}, K. 2014, \apj, 790, 42,
  \dodoi{10.1088/0004-637X/790/1/42}

\bibitem[{{Spalding} \& {Millholland}(2020)}]{Spalding2020}
{Spalding}, C., \& {Millholland}, S.~C. 2020, \aj, 160, 105,
  \dodoi{10.3847/1538-3881/aba629}

\bibitem[{{Van Eylen} \& {Albrecht}(2015)}]{Van2015}
{Van Eylen}, V., \& {Albrecht}, S. 2015, \apj, 808, 126,
  \dodoi{10.1088/0004-637X/808/2/126}

\bibitem[{{Vandakurov}(1967)}]{Vandakurov1967}
{Vandakurov}, Y.~V. 1967, \azh, 44, 786

\bibitem[{{Vogt} {et~al.}(1994){Vogt}, {Allen}, {Bigelow}, {Bresee}, {Brown},
  {Cantrall}, {Conrad}, {Couture}, {Delaney}, {Epps}, {Hilyard}, {Hilyard},
  {Horn}, {Jern}, {Kanto}, {Keane}, {Kibrick}, {Lewis}, {Osborne},
  {Pardeilhan}, {Pfister}, {Ricketts}, {Robinson}, {Stover}, {Tucker}, {Ward},
  \& {Wei}}]{Vogt1994}
{Vogt}, S.~S., {Allen}, S.~L., {Bigelow}, B.~C., {et~al.} 1994, Society of
  Photo-Optical Instrumentation Engineers (SPIE) Conference Series, Vol. 2198,
  {HIRES: the high-resolution echelle spectrometer on the Keck 10-m Telescope},
  362, \dodoi{10.1117/12.176725}

\bibitem[{{Weiss} \& {Marcy}(2014)}]{WM2014}
{Weiss}, L.~M., \& {Marcy}, G.~W. 2014, \apjl, 783, L6,
  \dodoi{10.1088/2041-8205/783/1/L6}

\bibitem[{{Winn} {et~al.}(2005){Winn}, {Noyes}, {Holman}, {Charbonneau},
  {Ohta}, {Taruya}, {Suto}, {Narita}, {Turner}, {Johnson}, {Marcy}, {Butler},
  \& {Vogt}}]{Winn2005}
{Winn}, J.~N., {Noyes}, R.~W., {Holman}, M.~J., {et~al.} 2005, \apj, 631, 1215,
  \dodoi{10.1086/432571}

\bibitem[{{Winn} {et~al.}(2010){Winn}, {Johnson}, {Howard}, {Marcy},
  {Isaacson}, {Shporer}, {Bakos}, {Hartman}, \& {Albrecht}}]{Winn2010}
{Winn}, J.~N., {Johnson}, J.~A., {Howard}, A.~W., {et~al.} 2010, \apjl, 723,
  L223, \dodoi{10.1088/2041-8205/723/2/L223}

\bibitem[{{Xie}(2014)}]{Xie}
{Xie}, J.-W. 2014, \apjs, 210, 25, \dodoi{10.1088/0067-0049/210/2/25}

\bibitem[{{Xuan} \& {Wyatt}(2020)}]{Xuan2020}
{Xuan}, J.~W., \& {Wyatt}, M.~C. 2020, \mnras, \dodoi{10.1093/mnras/staa2033}

\bibitem[{{Yee} {et~al.}(2018){Yee}, {Petigura}, {Fulton}, {Knutson},
  {Batygin}, {Bakos}, {Hartman}, {Hirsch}, {Howard}, {Isaacson}, {Kosiarek},
  {Sinukoff}, \& {Weiss}}]{Yee2018}
{Yee}, S.~W., {Petigura}, E.~A., {Fulton}, B.~J., {et~al.} 2018, \aj, 155, 255,
  \dodoi{10.3847/1538-3881/aabfec}

\bibitem[{{Zanazzi} \& {Lai}(2018)}]{zanazzi2018}
{Zanazzi}, J.~J., \& {Lai}, D. 2018, \mnras, 477, 5207,
  \dodoi{10.1093/mnras/sty951}

\bibitem[{{Zeng} \& {Seager}(2008)}]{zeng2008}
{Zeng}, L., \& {Seager}, S. 2008, \pasp, 120, 983, \dodoi{10.1086/591807}

\end{thebibliography}
\bibliographystyle{aasjournal}


\end{CJK*}
\end{document}